\documentclass[aps,pra,reprint,groupedaddress]{revtex4-1}
\usepackage{amsmath}
\usepackage{amssymb}
\usepackage{natbib}
\usepackage{graphicx}
\usepackage{color}
\newcommand{\eins}{\mbox{$1 \hspace{-1.0mm}  {\bf l}$}}
\newcommand{\psib}{\boldsymbol\psi}
\newcommand{\phib}{\boldsymbol\phi}
\newcommand{\etab}{\boldsymbol\eta}
\newcommand{\omb}{\boldsymbol\omega}

\newcommand{\Jb}{{\bf J}}

\newcommand{\nb}{{\bf n}}
\newcommand{\mb}{{\bf m}}
\newcommand{\Spin}{{\bf \hat{S}}}

\newcommand{\Ucal}{\mathcal{U}}

\newcommand{\Ncal}{\mathcal{N}}
\newcommand{\Gcal}{\mathcal{G}}

\newcommand{\Hcal}{\mathcal{H}}

\newcommand{\Hcalt}{\widetilde{\mathcal{H}}}

\newcommand{\Fcal}{\mathcal{F}}
\newcommand{\Acal}{\mathcal{A}}
\newcommand{\Tcal}{\mathcal{T}}

\newcommand{\Xcal}{\mathcal{X}}
\newcommand{\Ycal}{\mathcal{Y}}

\newcommand{\Ei}{{\bf E}}
\newcommand{\Di}{{\bf D}}
\newcommand{\ei}{{\bf e}}

\newcommand{\Om}{\boldsymbol\Omega}
\newcommand{\tpartial}{\widetilde{\partial}}
\newcommand{\tOm}{\widetilde{\Om}}
\newcommand{\tEi}{\widetilde{\Ei}}
\newcommand{\tDi}{\widetilde{\Di}}

\begin{document}

%\preprint{}

\title{Control of noisy quantum systems: Field theory approach to error mitigation}
%\title{Error mitigation schemes for the control of noisy quantum systems \\
%via Martin-Siggia-Rose and Schwinger-Keldysh field theory techniques}


\author{Rafael Hipolito and Paul M.~Goldbart}
%\thanks{}
\affiliation{School of Physics, Georgia Institute of Technology, 837 State Street, Atlanta, Georgia 30332 }


\date{\today}

\begin{abstract}
We consider the basic quantum-control task of obtaining a target unitary operation (i.e., a quantum gate) via control fields that couple to the quantum system and are chosen to best mitigate errors resulting from time-dependent noise, which frustrate this task.
We allow for two sources of noise: fluctuations in the control fields and fluctuations arising from the environment. We address the issue of control-error mitigation by means of a formulation rooted in the Martin-Siggia-Rose (MSR) approach to noisy, classical statistical-mechanical systems.  To do this, we express the noisy control problem in terms of a path integral, and integrate out the noise to arrive at an effective, noise-free description. We characterize the degree of success in error mitigation via a fidelity metric, which characterizes the proximity of the sought-after evolution to ones that are achievable in the presence of noise. Error mitigation is then best accomplished by applying the optimal control fields, i.e., those that maximize the fidelity subject to any constraints obeyed by the control fields.
To make connection with MSR, we reformulate the fidelity in terms of a Schwinger-Keldysh (SK) path integral, with the added twist that the ``forward'' and ``backward'' branches of the time-contour are inequivalent with respect to the noise.  The present approach naturally and readily allows the incorporation of constraints on the control fields---a useful feature in practice, given that constraints feature in all real experiments.
% Finally, we calculate the full probability distribution of the fidelity, allowing one to address questions of extreme value statistics, i.e. the likelihood of rare events.
 We illustrate this MSR-SK reformulation by considering a model system consisting of a single spin-$s$ freedom (with $s$ arbitrary), focusing on the case of $1/f$ noise in the weak-noise limit. We discover that optimal error-mitigation is accomplished via a universal control field protocol that is valid for all $s$, from the qubit (i.e., $s=1/2$) case to the classical (i.e., $s \rightarrow \infty$) limit. In principle, this MSR-SK approach provides a transparent framework for addressing quantum control in the presence of noise for systems of arbitrary complexity. 

\end{abstract}

%\keywords{}

\maketitle


%**************************************************************************************************************

\section{Introduction}

The ability to control the fate of quantum systems whose dynamics are subject to noisy influences is a critical factor in numerous settings, including, notably, quantum information processing~\cite{nielsen2000,domenico2007,wiseman2010}. 
A typical characteristic of such systems is that in order to \emph{control\/} them it is necessary to couple them to external time-varying fields that cannot themeselves be perfectly controlled and which, therefore, inevitably introduce classical noise into the system.
This noise, present in the external fields,  
alters the time-evolution of the quantum system of interest (i.e., the system we wish to control), typically pushing the end result away from the intended target. The external control fields enter the Hamiltonian governing the dynamics of the quantum system by way of coupling directly to the system's degrees of freedom, and thus coupling these sources of noise directly to it.  The dynamics, including the contribution from the noisy external fields, can be described via a Hamiltonian containing external \emph{stochastic\/} parameters in addition to terms describing internal system dynamics.   Although the probabilistic properties of the stochastic parameters can be determined, it is generally impossible to predict the values realized in any instance of the control attempt.
Even in the absence of these external control fields,
quantum systems are still generally subject to external sources of noise, because no system is truly isolated from the environment.
Environmental noise is detrimental to the control mission because it inevitably leads to the unremediable loss of information from the quantum system, due to entanglement between system and environment degrees of freedom, leading to quantum decoherence~\cite{bennett2000}. 

In the light of these remarks it is of value to identify and understand how to design schemes capable of mitigating the effects of noise to the greatest extent possible.
Several approaches have been developed to address this issue, including dynamical decoupling~\cite{viola1999}, dynamical control by modulation~\cite{kofman2001,kofman2004} and, more recently, the filter function approach used in Refs.~\cite{green2012,kabytayev2014}).
The aim of the present Paper is to develop an alternative approach to the task of quantum control in the presence of noise. The approach is rooted in the Schwinger-Keldysh (henceforth SK) path-integral framework~\cite{schwinger1960,keldysh1965}, which has frequently
been invoked in treatments of quantum dynamical systems that are not in thermodynamic equilibrium; see, e.g., Refs.~\cite{kamenev2011,polkovnikov2010}.
The SK framework is especially well suited for providing a transparent account of the effects of quantum fluctuations (i.e., \emph{quantum noise}) via interference between quantum fields that propagate `forward' and `backward' in time.  Although the framework was originally developed with \emph{closed} quantum systems in mind, it can readily be modified to account for open quantum systems that are coupled to external fields and sources of noise.

The core idea behind the present approach is as follows. 
We consider noisy quantum systems for which we possess:
(i)~a complete characterization of the noise-free system via the specification of its freedoms and a Hamiltonian that governs them; and 
(ii)~a complete \emph{statistical} characterization of the noise that perturbs the dynamics from its noise-free form, in the form a probability distribution for the history of the noise parameters. We take the goal of the control process to be to guide the system as accurately as possible, i.e., to impart upon it (as accurately as possible) some predetermined unitary transformation, 
without being informed about the instance of the noise-parameter history. 
We characterize the performance of the control process---i.e., its ability to impart a predetermined target operation upon the quantum system of interest---via an overlap metric or \emph{fidelity\/}, which is designed to assess the accuracy of the process, averaged over the noise-parameter history weighted by its known distribution. 
The guiding of the system is accomplished via a time-dependent Hamiltonian that we select from some menu; typically the menu is incomplete, in the sense that only certain operators are regarded as being available and the time-dependence of the classical variables that characterize these operators is restricted by the kinds of constraints present in real experiments. 
An inevitable consequence of the present framework is that the guiding that will be ascertained will be more accurate for some noise histories and target unitaries than for others. 
A strength, however, is that it need only be determined once, for any given noisy system and noise distribution.  

We formulate the task of controlling a generic noisy quantum system via an \emph{optimization\/} problem, in which one seeks the control-field history that maximizes the fidelity, i.e., the measure of success alluded to earlier; see also Refs.~\cite{green2012,kabytayev2014} (although the present definition of the fidelity differs slightly from those used in these references).
The path-integral formulation of the fidelity (which is the object of primary interest to us) takes a form that is close to the conventional SK path integral but has the following twist: The \lq forward-in-time\rq\ and \lq backward-in-time\rq\ branches are asymmetrical with respect to the \emph{external\/} noise (both environmental and due to the controls), which is present in the former branch but absent from the latter. 
This is a direct consequence of the definition we have chosen for the fidelity. By contrast, the internal \emph{quantum noise}, which originates within the system of interest still appears symmetrically in the two branches, as it does in the usual SK path integral. 
The SK formulation of control is similar in spirit to the approach introduced by Martin-Siggia-Rose (MSR)~\cite{msr1973} to study classical statistical dynamics.  % 
Once we have constructed the path integral for our SK-type formulation of the fidelity, we integrate out the environmental noise and thus arrive at an effective description that is completely deterministic (i.e., noise free), to which we can apply the tools available from field theory, such as diagrammatic expansion and even nonperturbative methods. 	

Via our approach, we show that the optimization procedure used to maximize the fidelity naturally gives rise to an action principle, which leads to equations of motion for the control-field history. The solution of these equations is a continuous deformation of the control scheme relevant to the noise-free case.  As one continuously increases the strength of the noise, the optimal control scheme continuously deforms, parametrically, away from the noise-free scheme.  The noise-free scheme typically presents an arbitrary number of schemes to choose from.  By adopting any one of these schemes and continuously increasing the strength of the noise, we sweep through a continuous \emph{family} of control schemes, with members of the family being parametrized by the noise strength.  Different choices of noise-free schemes typically correspond to different sets of winding numbers, as we show below.  
Each family can then be labeled by its set of winding numbers, which are invariant under these continuous deformations.  
%Since the optimization procedure gives rise to equations of motion derived from an action principle,  one can harness physical intuition %from action principles in 
%and has access to tools available in the study of classical systems .%to aid in the interpretation of these solutions.  Our approach is in some sense complementary to the use of pulse sequences, because we work with continuous deformations of continuous control functions as opposed to sequences of idealized sharp pulses. 
In common with more familiar action principles, the present approach has the advantage of allowing the straightforward implementation of many classes of constraints, including those naturally appearing in experiments, via the Lagrange multiplier technique.  
The path-integral formalism also provides a natural starting point for developing a semiclassical approach to the task of controlling  noisy quantum systems, in which quantum effects are introduced as a refinement of an underlying classical process. The literature is extensive on controlling noisy \emph{classical\/} systems, as it is on the control of two-level quantum systems (i.e., qubits), which constitute the extreme quantal case.  \emph{Inter alia}, the formalism we develop here serves to bridge the gap between these two extremes---a regime that has been left relatively unexplored, to date.

We illustrate our approach by analyzing some concrete examples in detail.  We specifically develop the formalism for a single quantum spin $\hat{{\bf S}}$, keeping the spin quantum number $s$ arbitrary.  We couple the spin to external control fields in order to achieve a target operation, and allow the spin to be under the influence of noise (from the environment or the control fields or both) with the statistics of the noise presumed known. Although we develop the formalism with this specific system in mind, we note that it can be readily generalized to more elaborate systems, including interacting systems such as spin chains and ultracold atomic gases.  We also note that the SK path integral, if formulated in terms of coherent states, provides a natural starting point for a semiclassical expansion.  This is a useful feature, if one is interested in studying control problems for systems where the semiclassical expansion gives an excellent approximation to the full quantum dynamics (e.g., ultracold bosonic gases~\cite{polkovnikov2011}).

Continuing with the case of a single spin, we use our formalism to find the fidelity and the corresponding optimized control fields for various choices of noise distributions, with a special focus on $1/f$ noise sources, which arise in many systems of interest~\cite{1freview}.  For the specific case of a single spin, we find results for arbitrary $s$, ranging from the qubit limit ($s=1/2$) up to the classical limit $s\to \infty$.  
%In addition to finding the fidelity itself, we will also use our formalism to compute the exact probability distribution for all quantities of interest in this control problem in the weak noise limit---which, to the knowledge of the authors has not been investigated, at least in recent literature.  This opens up the possibility of tackling questions concerning extreme value statistics, such as the likelihood of catastrophic events~\cite{fortin2015,bouchaud1997}.

The Paper is organized as follows. 
In Sec.~\ref{sec:theory} we give a discussion of the general setting that we shall be concerned with, 
and also review the various questions that we shall be exploring in detail.
In Sec.~\ref{sec:spins} we formulate the control problem in terms of a modifed SK approach, 
and apply it to the case of a single spin ${\bf \hat{S}}$ of arbitrary spin quantum number $s$ in the presence of noise.
We also construct the expression for the fidelity and find the optimal control scheme for the case of weak $1/f$ noise and  constraints being imposed on the strength of the control fields. 
%and find the full probability distribution for the fidelity in the case of weak noise.  
In Sec.~\ref{sec:conclusion} we summarize our results and briefly discuss future directions. 
The technical details for the derivation of our formalism are mostly relegated to appendices for the sake of the clarity of the presentation, and will be referred to where relevant.

%******************************************************************************************************************

\section{Elements} \label{sec:theory}
Our task is to control a given quantum system --- specifically, we wish to complete a \emph{predetermined unitary transformation}, which we call the \emph{target unitary operator\/} $U_T$, upon the quantum state. The state is not necessarily known ahead of time.  In order to accomplish this, we invoke external time-dependent \emph{control fields}, which couple directly to the system degrees of freedom and are used to steer the quantum system in such a way that, at the end of some transit time $\tau$, the quantum system has evolved to a final state that is equivalently described by $U_T$ in the sense that the \emph{net effect} of the two is the same.  

The quantum system we wish to control is not generally isolated. It is coupled to an external environment and is thus subject to environmental noise.  In addition, there may be fluctuations in the control fields themselves (i.e., originating in the devices used to generate these fields), which provide another source of noise. The net effect originating from all sources of noise will generally interfere with the control scheme. Assuming one chooses the control fields such that one obtains a unitary equivalent to $U_T$ at the end of the transit time $\tau$ \emph{in the absense of noise}, these same control fields will generally give rise to a unitary transformation that differs from $U_T$.  

The question we address can be stated as follows: given many histories of \emph{sets} of control fields to choose from, each of which would give rise to $U_T$ in the absence of noise, which particular \emph{set} gives rise to a unitary transformation that is \emph{closest\/} to $U_T$? In the presence of noise, it is not possible to predict how the quantum system that we aim to control will evolve in each given instance of the noise field history. The best we can do is determine with a \emph{statistical} measure that tells us how close we are to reaching our stated goal, which is to have unitary evolution that is as close as possible to $U_T$ at the end of the transit time $\tau$.  

The simplest such measure, viz., the \emph{fidelity} (which we define below), reports how close we get \emph{on average\/}, i.e., after averaging over all sources of noise.  In other words, we choose a \emph{single} set of control fields; then, for each instance of noise, we find the corresponding unitary transformation at the end of the transit time $\tau$; we \emph{average} these unitary transformations over all instances of noise histories; and, finally, we compare the resulting averaged unitary transformation with $U_T$. We define the best single set of control field histories --- what we are searching for --- to be the set from which one obtains the (averaged) transformation that lies as close as possible to the one resulting from $U_T$.  We rigorously define all of the quantities of interest below, as we develop our methodology.     

Before developing our methodology, let us spend some time to obtain a better quantitative understanding of the problem at hand  in terms of the various Hamiltonian terms involved.   

%**********************

\subsection{Hamiltonian}
Consider a quantum system described by a Hamiltonian $H(t)$, which can be written as a sum of several parts:
\begin{eqnarray}
H(t) &=& H_c(t) + H_s(t) + H_e(t) + \epsilon_{se} H_{se}(t); \label{eq:Htotal}
\end{eqnarray}
we allow all parts of the Hamiltonian to be explicitly time dependent, and $H_c(t)$, $H_s(t)$, $H_e(t)$, and $H_{se}(t)$ are respectively taken to be the control Hamiltonian, system Hamiltonian, environmental Hamiltonian, and system-environment coupling Hamiltonian.   The parameter $\epsilon_{se}$ determines the strength of the system-environment coupling, which we take to be weak.  

Let us denote by $\{\hat{q}_i\}$ the quantum degrees of freedom of the system (i.e., the part that we wish to \emph{control\/}), and by  $\hat{Q}_i$ the degrees of freedom of the environment. The various Hamiltonian terms have the following characteristics: $H_c(t)=\sum_i c_i(t)\,\hat{q}_i$ couples the \emph{control fields} $c_i(t)$ directly to the quantum degrees of freedom $\hat{q}_i$, which we are interested in controlling; $H_s(t)=H_s(t,\hat{q}_i)$ determines the internal system dynamics; $H_{e}(t)=H_{e}(t,\hat{Q}_i)$ determines the dynamics of the environment; and $H_{se}(t)=H_{se}(t,\hat{q}_i,\hat{Q}_i)$ couples the system and environment to one another.  From the perspective of RG, the terms that linearly couple the system and environment degrees of freedom are the most relevant (see, e.g. Ref.~\cite{weiss1999}).  Taking the system-environment coupling to be weak (i.e., $\epsilon_{se}$ small), and temporarily ignoring all other terms in $H_{se}(t)$ we get
\begin{eqnarray}
H_{se}(t) &\cong&\sum_{ij} \hat{q}_i\,\eta_{ij}(t)\,\hat{Q}_j. 
\end{eqnarray}
Next, by using the path integral language we can proceed to \emph{integrate out} the environment degrees of freedom, and thus obtain an effective Hamiltonian, in which only the system degrees of freedom remain.  Carrying out this procedure, and assuming that one can treat the environmental degrees of freedom using a semiclassical approximation, %(i.e. the environment is not too close to its ground state),
we obtain an effective system-environment Hamiltonian $\widetilde{H}_{se}(t)$ which in general has the following form:
\begin{eqnarray}
\widetilde{H}_{se}(t) &\simeq& \sum_i \tilde{\eta}_i(t)\,\hat{q}_i,
\end{eqnarray}
where the histories $\{\tilde{\eta}_i(\cdot)\}$ are \emph{stochastic}, with a histories functional probability density $P_e[\{\tilde{\eta}_i(\cdot)\}]$ (where $\{\tilde{\eta}_i(\cdot)\}$ designates the set of $\tilde{\eta}_i(\cdot)$ for all $i$), which in principle can be determined from the state of the environment and the environment Hamiltonian $H_e$.  

In actual experiments, there is also some degree of uncertainty associated with the control fields $c_i(t)$ themselves, such as fluctuations in the fields due to noise generated in the experimental apparatus responsible for the generation of these fields. In general, we can write
\begin{eqnarray}
c_i(t) &=& \tilde{c}_i(t)+ \epsilon_c \,\delta c_i(t),
\end{eqnarray} %%%%%%%%%%%%%%%%%%%%%
where $\tilde{c}_i(t)$ is the control field as given in the absence of any fluctuations, the fluctuations $\delta c_i(t)$ are described by some probability distribution $P_c(\{\delta c_i(\cdot)\})$, and $\epsilon_{c}$ is a parameter describing the strength of fluctuations.  As experimental measurements are sensitive to the \emph{net\/} contribution of noise, we combine the environmental noise and noise due to fluctuations in the control fields into a single effective noise term, $H_n(t)$, given by
\begin{eqnarray}
\epsilon H_n(t) &=& \epsilon \sum_i n_i(t)\,\hat{q}_i
\end{eqnarray}
where $\epsilon\equiv\epsilon_{se}+\epsilon_{c}$ and
\begin{eqnarray}
n_i(t) &\equiv& \frac{\epsilon_{se}}{\epsilon_{se}+\epsilon_{c}} \tilde{\eta}_i(t) + \frac{\epsilon_{c}}{\epsilon_{se}+\epsilon_{c}}\delta c_i(t), 
\end{eqnarray}
so that $\left\{n_i(\cdot)\right\}$ are effective stochastic fields described by a (classical) effective probability distribution $P[\{n_i(\cdot)\}]$, which can in principle be determined via experiment.  In what follows, we assume that we know $P[\{n_i(\cdot)\}]$. 

Let us replace $\tilde{c}_i$ by $c_i$, where $c_i$ is now understood to be the control field.  We can now specify the effective Hamiltonian, which we will refer to as $H_{\epsilon}(t)$, so as to remind us of its dependence on $\epsilon$ [note this Hamiltonian is different from Eq.~(\ref{eq:Htotal})]
\begin{eqnarray}
H_{\epsilon}(t) &=& H_s(t,\hat{q}_i) + \sum_i\big( c_i(t) + \epsilon n_i(t)  \big)\hat{q}_i \nonumber\\
 &\equiv& H_s(t) + H_c(t) + \epsilon H_n(t). \label{eq:Ht}
\end{eqnarray}
This effective Hamiltonian, $H_{\epsilon}$, contains the full quantum description of the system degrees of freedom $\hat{q}_i$, which are coupled to both  the \emph{stochastic} fields $n_i(t)$ and the \emph{control} fields $c_i(t)$, as well as to one another via the internal system dynamics as described by $H_s$.

%****************

\subsection{Fidelity}
We now construct the \emph{fidelity}, i.e., a measure of success in effecting the specified unitary transformations, averaged over the noise history.   At the end of the transit time $\tau$, our aim is to have the system evolution to be as close as possible to $U_T$.  The evolution operator $U_{\epsilon}(t,\{c_i(\cdot)\},\{n_i(\cdot)\})$ corresponding to $H_{\epsilon}$, cf.~Eq.~(\ref{eq:Ht}), is given by
\begin{eqnarray}
U_{\epsilon}(\tau,\{c_i(\cdot)\},\{n_i(\cdot)\})&=&\mathcal{T}\exp\left[-i\int_0^{\tau}\mathrm{d}t \,H_{\epsilon}(t)  \right] \nonumber\\
\end{eqnarray}
 where the notation reminds us that $U_{\epsilon}$ is a function of $\tau$ as well as a \emph{functional\/} of the control and effective noise fields.  We shall often use the shorthand $U_{\epsilon}(t)$, provided there is no risk of confusion.  The quantity $U_0(t)\equiv U_{\epsilon}(t)|_{\epsilon=0}$ corresponds to the evolution operator in the absence of noise.  Recall that at the transit time $\tau$, we have $U_0(\tau)=U_T$ (i.e. the target unitary).  We elect to define the fidelity $\Fcal$ as follows:
\begin{subequations}
\begin{eqnarray}
\Fcal(\{c_i(\cdot)\}) &=& \langle\,\mathrm{Tr}\, U_T^{\dag}\, U_{\epsilon}(\tau) \,\rangle_n \\
&=&  \langle \,\mathrm{Tr}\, U_0^{\dag}\, (\tau) U_{\epsilon}(\tau)  \, \rangle_n \label{eq:fidel}
\end{eqnarray}
\end{subequations}
where the brackets $\langle \cdots \rangle_n$ denote averaging over $\{n_i(\cdot)\}$, the trace operation is taken over the entire Hilbert space of the system, normalized by the dimension of the Hilbert space so that $\mathrm{Tr}\,\eins =1 $.  Observe that the fidelity obeys $0\leq |\Fcal| \leq 1$ and reaches its upper bound of unity if and only if $U_{\epsilon}=U_T$ (i.e., perfect control is achieved regardless of initial state).
The fidelity is a \emph{functional} of the control fields $\left\{c_i(\cdot)\right\}$. In order to best accomplish the sought for unitary transformation, we solve the \emph{variational} problem to find the set $\{c_i(\cdot)\}$ that \emph{maximizes} $\Fcal$.  We shall make use of the formulation given in the second line of Eq.~(\ref{eq:fidel}), which turns out to be efficacious when expressed in terms of a path integral.  In terms of the Hamiltonian, the fidelity is given by
\begin{eqnarray}
\Fcal[\{c_i(\cdot)\}] &=& \Big\langle  \mathrm{Tr}\, \mathcal{T}_K \,\mathrm{e}^{i\int_0^{\tau}\mathrm{d}t\,H_0(t)}\,\mathrm{e}^{-i\int_0^{\tau}\mathrm{d}t\, H_{\epsilon}(t) }\Big\rangle_n. \label{eq:fidelpi}
\end{eqnarray}
Here the operation $\mathcal{T}_K$ corresponds to time-ordering on the Keldysh contour (see, e.g., Ref.~\cite{kamenev2011}): the first exponential factor is \emph{anti\/}-time ordered; the second one is time-ordered in the usual way.  Equation~(\ref{eq:fidelpi}) can be expressed in terms of a Schwinger-Keldysh path integral over a closed-time contour, but there is a twist: unlike the mere usual quantum-dynamical problems (e.g., for isolated quantum systems), for which this technique is commonly applied, in the present setting there is an \emph{asymmetry} between the forward and backward branches of the time contour.  Specifically, the stochastic degrees of freedom $\{n_i(\cdot)\}$ are only present in the forward branch.  

As we shall see, for some purposes, it is useful to refer to the \emph{fidelity amplitude} $\Acal$ instead of the fidelity.  This is defined via
\begin{eqnarray}
\Acal[\{c_i(\cdot)\},\{n_i(\cdot)\}] &=&  \mathrm{Tr}\, \mathcal{T}_K \,\mathrm{e}^{i\int_0^{\tau}\mathrm{d}t\,H_0(t)}\,\mathrm{e}^{-i\int_0^{\tau}\mathrm{d}t\, H_{\epsilon}(t) }, \nonumber\\
&&  \label{eq:fidamp}
\end{eqnarray} 
which is just the expression for the fidelity \emph{before\/} taking the average over the stochastic fields, and thus is a functional of \emph{both} $\{c_i(\cdot)\}$ and $\{n_i(\cdot)\}$.  In terms of $\Acal$ we have
\begin{eqnarray}
\Fcal[\{c_i(\cdot)\}] &=& \langle \Acal(\{c_i(\cdot)\},\{n_i(\cdot)\}) \rangle_n
\end{eqnarray}

%***********************

\subsection{Constraints}
Determining the control sequence that best mitigates noise amounts to finding the set of control fields $\{c_i(\cdot)\}$ that maximize $\Fcal$  and yet satisfy all \emph{constraints\/} imposed on the set $\{c_i(\cdot)\}$ \emph{together with\/} the boundary condition at the end of the transit time $\tau$, viz.~$U_0(\tau)=U_T$.  In any realistic situation, there will be physical limitations on $\{c_i(\cdot)\}$ that we may consider. For instance, each $c_i(t)$ must be of \emph{finite magnitude\/}, and its functional dependence on time would be constrained by the experimental devices in use.  

There are certain classes of constraints that can be entirely accounted for by using Lagrange multipliers. These include, but are not limited to, holonomic constraints.  Explicit examples will be worked out in the following sections.  

\section{Single spin ${\bf \hat{S}}$} \label{sec:spins}
We now spell out explicitly how the abstract formalism presented in Sec.~\ref{sec:theory} applies in the case of a single spin $\Spin$, for now leaving the spin quantum number $s$ arbitrary.  In this setting, the total Hamiltonian is given by
\begin{eqnarray}
H_{\epsilon}(t) &=& (\omb(t)+\epsilon {\bf n}(t))\cdot \Spin 
\end{eqnarray}
where the field $\omb(t)$ corresponds to the external control field and ${\bf n}(t)$ is the stochastic field, discussed in Sec.~\ref{sec:theory},  that represents the effect of environmental noise and also accounts for any fluctuations in the control field inherently present in the devices used to generate it. In addition, $\omb(t)$ is the control in the absence of such fluctuations.  In the present example, the system Hamiltonian,  $H_s(t)$, vanishes.  

Our goal, then, is to determine the $\omb(t)$ that maximizes the fidelity $\Fcal$ at the end of the transit time $\tau$.  In other words, we seek $\omb(t)$ such that, after averaging over different realizations of ${\bf n}(t)$, the evolution operator is as close as possible to some prescribed target $U_T$.  

As discussed in Sec.~\ref{sec:theory}, the stochastic fields $\{n_i(\cdot)\}$ are governed by a distribution functional $P[\{n_i(\cdot)\}$ which is presumed to be known.  In the present case, we take $P$ to be Gaussian with zero mean and covariance given by
\begin{eqnarray}
\langle n_i(t)n_j(t) \rangle_n &=& \Ncal_{ij}(t,t'). \label{eq:ninj}
\end{eqnarray}
We restrict our attention to stochastic fields that are stationary in time and time-reversal invariant, in which case
\begin{equation}
\Ncal_{ij}(t,t') =\Ncal_{ij}(|t-t'|) = \Ncal_{ji}(|t-t'|). 
\end{equation}

We make use of the Schwinger-Keldysh (SK) path-integral formulation (see, e.g.~Refs.~\cite{schwinger1960,keldysh1965} for original work by Schwinger and Keldysh and Ref.~\cite{kamenev2011} for a modern treatment), which for the present system can be done either in terms of spin coherent states or bosonic coherent states (if one makes use of the Schwinger representation for spins), see e.g.~Ref.~\cite{klauder1985}.   Choosing the latter, we make use of the mapping between the spin operator $\Spin$ and the two-component bosons ${\bf \hat{a}}^{\dag}\equiv (a_1^{\dag}, a_2^{\dag})$, given by
\begin{eqnarray}
\Spin &=& \frac{1}{2} {\bf \hat{a}}^{\dag}\cdot \boldsymbol\sigma \cdot {\bf \hat{a}}
\end{eqnarray}
where $\boldsymbol\sigma=(\sigma_x,\sigma_y,\sigma_z)$ are the Pauli matrices.  The total spin quantum number $s$ is conserved by the dynamics, and is related to the total number of bosons in the Schwinger representation: 
$s = \frac{1}{2} {\bf a}^{\dag}\cdot {\bf a}$.
The total Hamiltonian then takes the form:
\begin{eqnarray}
H_{\epsilon}(t) &=& {\bf a}^{\dag} \cdot \Hcal_{\epsilon}(t)\cdot {\bf a},
\end{eqnarray}
where 
\begin{eqnarray}
\Hcal_{\epsilon} &=& \frac{1}{2}\omb(t)\cdot\boldsymbol\sigma+\frac{\epsilon}{2} {\bf n}(t) \cdot \boldsymbol\sigma \nonumber\\
&\equiv& \Hcal_c(t) + \epsilon\Hcal_n(t) .
\end{eqnarray}

%***************

\subsection{Schwinger-Keldysh (SK) path integral}

We now construct the path-integral expression for the fidelity amplitude $\Acal_s$, Eq.~(\ref{eq:fidamp}), where $s$ indicates the (arbitrary) spin quantum number.  We make use of the coherent state basis $| \boldsymbol\alpha\rangle\equiv |\alpha_1, \alpha_2 \rangle$, for complex $\alpha_1$ and $\alpha_2$, defined by ${\bf a}|\boldsymbol\alpha \rangle=\boldsymbol\alpha | \boldsymbol\alpha \rangle$ (see Ref.~\cite{klauder1985}), in order to evaluate the SK path integral along the SK contour. 
We use the labels $\boldsymbol\alpha_{f,b}$ for the forward-in-time and backward-in-time branches along the contour, respectively.  We thus obtain the expression
\begin{eqnarray}
\Acal_s &=& \mathrm{Tr}\int \textrm{D}^2\boldsymbol\alpha_f(t)\textrm{D}^2\boldsymbol\alpha_b(t)\times|\boldsymbol\alpha_b(0)\rangle\nonumber\\
&& \times \mathrm{e}^{-\frac{1}{2}[|\boldsymbol\alpha_b(\tau)|^2+|\boldsymbol\alpha_f(0)|^2]+\boldsymbol\alpha_b^{\star}(\tau)\cdot \boldsymbol\alpha_f(\tau)} \nonumber\\
&&\times \mathrm{e}^{i\int_0^{\tau}\mathrm{d}t\, [\boldsymbol\alpha_f^{\star}\cdot (i\partial_t-\mathcal{H}_{\epsilon}) \cdot \boldsymbol\alpha_f-\boldsymbol\alpha_b^{\star}\cdot(i\partial_t-\mathcal{H}_c) \cdot \boldsymbol\alpha_b]}\langle\boldsymbol\alpha_f(0)| ,\nonumber\\
&& \label{eq:acali}
\end{eqnarray}
where the normalized trace $\mathrm{Tr}$ is taken over the complete set of two-mode bosonic number states $|n, 2s-n \rangle_{\rm num}$ for $n=0,1,\cdots,2s$, satisfying the constraint $\langle n,2s-n|\frac{1}{2}{\bf \hat{a}}^{\dag}\cdot {\bf \hat{a}}  |n,2s-n \rangle_{\rm num} = s $, so as to fix the quantum spin number $s$.  For further details on the meaning of Eq.~(\ref{eq:acali}), as well as explicit expressions for the measures $\textrm{D}^2\boldsymbol\alpha_{b,f}$ and other details concerning the path integral, we refer the reader to App.~\ref{app:pathint}.
Note that our approach differs from conventional SK approach~\cite{kamenev2011} used to study nonequilibrium quantum dynamics, in that the forward and backward branches of the time evolution are asymmetric with respect to noise: it is present in the forward branch and completely absent in the backward branch. 

The expression for $A_s$ in Eq.~(\ref{eq:acali}) can be evaluated with the help of a generating functional $G_s[\Jb]$, which we define as follows:
\begin{eqnarray}
G_s[\Jb] &=&  \mathrm{Tr}\int \textrm{D}^2\boldsymbol\alpha_f(t)\textrm{D}^2\boldsymbol\alpha_b(t)\times|\boldsymbol\alpha_b(0)\rangle\langle\boldsymbol\alpha_f(0)| \nonumber\\
&& \times \mathrm{e}^{-\frac{1}{2}[|\boldsymbol\alpha_b(\tau)|^2+|\boldsymbol\alpha_f(0)|^2]+\boldsymbol\alpha_b^{\star}(\tau)\cdot \boldsymbol\alpha_f(\tau)} \nonumber\\
&&\times \mathrm{e}^{i\int_0^{\tau}\mathrm{d}t \,[\boldsymbol\alpha_f^{\star}\cdot (i\partial_t-\mathcal{H}_c) \cdot \boldsymbol\alpha_f-\boldsymbol\alpha_b^{\star}\cdot(i\partial_t-\mathcal{H}_c) \cdot \boldsymbol\alpha_b]} \nonumber\\
&& \times \mathrm{e}^{\int_0^{\tau}\mathrm{d}t \,[\Jb^{\star}\cdot\boldsymbol\alpha_f + \boldsymbol\alpha_f\cdot\Jb]}, \label{eq:gjb}
\end{eqnarray}
where, as anticipated, this expression contains the \emph{noise-free} Hamiltonian on both branches of the Keldysh contour. Moreover, we have coupled the two-component external source $\Jb(t)$ to the forward branch only. This allows us to calculate quantum averages involving the forward branch fields. Recall that noise couples exclusively to this branch. Observe that we have $G_s[0]=1$ since in the absence of a source, $G_s$ is just the usual SK partition function and the asymmetry due to the noise no longer arises.  Physically, this corresponds to the feature that in the absence of noise the fidelity is exactly unity, as is natural.

In terms of $G_s[\Jb]$, we may evaluate the fidelity amplitude as follows:
\begin{eqnarray}
\Acal_s &=& \mathrm{e}^{-i\epsilon\int_0^{\tau}\mathrm{d}t \; \frac{\delta}{\delta\Jb(t)}\cdot \mathcal{H}_n(t) \cdot \frac{\delta}{\delta\Jb^{\star}(t)}} G_s[\Jb(t)]\Bigg|_{\Jb(t)={\bf 0}}. \nonumber\\
&& \label{eq:acalder}
\end{eqnarray}
%Here $\mathcal{H}_n(t)$ corresponds to the noise term in the Hamiltonian for the spin $s=1/2$ case.
$G_s[\Jb]$ is evaluated in App.~\ref{app:pathint}, giving 
\begin{eqnarray}
G_s[\Jb] &=& \oint\frac{\mathrm{d}z}{2\pi i}\frac{\mathrm{e}^{\int_0^{\tau}\int_0^{\tau}\mathrm{d}t\,\mathrm{d}t'\;\Jb^{\star}(t) \cdot \Gcal_c(z,t,t')\cdot \Jb(t') }}{(2s+1)(1-z)^2z^{2s+1}} \label{eq:gjbeval}
\end{eqnarray}
where the integral over the auxiliary complex variable $z$ is taken over any closed contour that encircles the origin once, but does not include the pole at $z=1$.  The variable $z$ plays the role of a conjugate variable to the discrete spin number $s$ --- this integral can be interpreted as an integral transform between the $z$ representation and $s$ representation.

In the exponent of Eq.~(\ref{eq:gjbeval}) we have the Green function $\Gcal_c(z,t,t')$, which is given by
\begin{eqnarray}
\Gcal_c(z,t,t')\equiv \left(\frac{z}{1-z}+\Theta(t-t')\right)\Ucal_c(t,t'),
\end{eqnarray}
where $\Theta(t)$ is the Heaviside step function (zero and unity for negative and positive arguments, respectively), and $\Ucal_c(t,t')\equiv\Tcal \exp[-i\int_{t'}^{t}\mathrm{d}t''\,\Hcal_c(t'')]$ is the unitary matrix corresponding to the control Hamiltonian. Note that in this expression and formula~(\ref{eq:acalder}) for $\Acal_s$, for an arbitrary spin $s$, we only need consider the dynamics of a two level system (i.e. spin $1/2$).  
%In practice, this is a useful simplification as far as questions of dynamics are concerned: one need only work with  $2\times2$ matrices no matter what the value of the quantum spin number $s$. 

To make the meaning (and evaluation) of $A_s$ more transparent, we can re-express Eq.~(\ref{eq:acalder}) in a terms of a two-component Gaussian quantum field $\phib(t)^{\dag}\equiv\big(\phi_1^{\star}(t),\phi_2^{\star}(t)\,\big)$ whose description is completely given in terms of the expectation value 
\begin{eqnarray}
\langle\phib(t) \, \phib^{\star}(t')  \rangle_q &=& \Gcal_c(z,t,t'),
\end{eqnarray}
where $\langle \cdots \rangle_q $ denotes a quantum average over $\phib(t)$, and higher-order averages are determined by means of Wick's theorem.  Thus,
\begin{eqnarray}
\Acal_s &=&  \oint\frac{\mathrm{d}z}{2\pi i} \frac{\left\langle\mathrm{e}^{-i\epsilon\int_0^{\tau}\mathrm{d}t\, \phib^{\star}\mathcal{H}_n \phib}   \right\rangle_q}{(2s+1)(1-z)^2 \, z^{2s+1}}. \label{eq:acalp}
\end{eqnarray}
Note the the expressions for $\Acal_s$ given in Eqs.~(\ref{eq:acalp}) and (\ref{eq:acalder}) are entirely equivalent.

The quantum expectation value for the fields $\phib(t)$ in Eq.~(\ref{eq:acalp}) can be evaluated exactly with the use of diagrammatic techniques, which are developed in Section 2 of App.~\ref{app:pathint}. Through an application of these techniques, we obtain the expression:
\begin{eqnarray}
\langle\mathrm{e}^{-i\epsilon \int_0^{\tau}\mathrm{d}t\;\phib^{\star}\mathcal{H}_n\phib } \rangle_q &=& \frac{(1-z)^2}{\mathrm{Det} [ \eins-z \Tcal \mathrm{e}^{-i\epsilon\int_0^{\tau}\mathrm{d}t\;\Ucal_c^{\dag}\mathcal{H}_n\Ucal_c}]}.\nonumber\\
&& \label{eq:qexpval}
\end{eqnarray}
Inserting this expression into Eq.~(\ref{eq:acalp}) and evaluating the integral over $z$, we obtain the result for the fidelity amplitude $\Acal_s$.  Recalling that $\Acal_s$ are functionals of the control $\omb(t)$ and the noise ${\bf n}(t)$, we finally obtain
\begin{eqnarray}
\Acal_s[\omb(\cdot),{\bf n}(\cdot)] &=& \frac{1}{2s+1}\sum_{j=-s}^s\mathrm{e}^{-2ji \cos^{-1}\big(\Acal_{1/2}[\omb(\cdot),{\bf n}(\cdot)]\big)},\nonumber\\
&& \label{eq:acalS}
\end{eqnarray}
where we are emphasizing the functional dependence of $\Acal_s$ on $\omb(\cdot)$ and ${\bf n}(\cdot)$. The quantity $\Acal_{1/2}[\omb(\cdot),{\bf n}(\cdot)]$, appearing in Eq.~(\ref{eq:acalS}), is the fidelity amplitude for the spin-half case. It is given by the expression
\begin{eqnarray}
\Acal_{1/2}[\omb(\cdot),{\bf n}(\cdot)] &=& \frac{1}{2}\mathrm{Tr}\, \Tcal \mathrm{e}^{-i\int_0^{\tau}\mathrm{d}t\;\Ucal_c^{\dag}(t)\,\Hcal_n(t) \, \Ucal_c(t)}\nonumber\\
&& \label{eq:ahalf}
\end{eqnarray}
in which $\Tcal$ denotes the usual time-ordering operator, $\tau$ is the transit time, and 
\begin{eqnarray}
\Ucal_c(t) &=& \Tcal \mathrm{e}^{-i\int_0^{t}\mathrm{d}t'\,\Hcal_c(t')} \label{eq:ucalc}
\end{eqnarray}
is the evolution operator corresponding to the control Hamiltonian.
The derivation of Eqs.~(\ref{eq:qexpval},\ref{eq:acalS}) is given in Section 2 of App.~\ref{app:pathint}.

%There are two remarkable properties regarding the result (\ref{eq:acalS}), which we briefly mention here.  First, one sees that the dynamical content for arbitrary spin $s$ is fully encoded in the spin-half formula.  Second, the resulting expression (\ref{eq:acalS}), an \emph{exact\/} quantum result, is equivalent to what one would obtain via a semiclassical calculation.   
%The first statement is a consequence of the latter -  for general systems, it is no longer true that semiclassical results are equivalent to exact quantum results.  Our methods, however, are applicable for all general systems - the diagrammatic methods are still there (as are nonperturbative methods), we just would simply not be able to sum up all diagrams to obtain the exact result as we have done here. In most cases of interest approximate results suffice and are straightforward to obtain using our methods. We note, however, that though we can obtain exact results, the results we get are nontrivial and the system we consider is interesting from the point of view of quantum control. 
% For purposes of physical clarity, however, there are advantages to working with a system where one can obtain exact results.

%We first obtain the expression for the fidelity amplitude $\Acal_s(\omb(t),{\bf n}(t))$ by evaluating the path integral (shown in the appendix) for the case of \emph{arbitrary} spin number $s$

%******************************

\subsection{Connection with quaternions}

Before continuing with our development, let us pause to streamline our notation.  To this end, we observe a connection with quaternions that makes the physics more transparent. The advantage stems from the fact that quaternions provide a unified method for treating both geometrical $3$-vectors and unitary evolution operators, all under the language of quaternion algebra.  In what comes below, the reader may freely replace pure quaternions (see App.~\ref{app:quat}) by vectors in all settings, except when they occur in exponents.  In that case, the correct interpretation is to replace pure quaternions by the corresponding vector dotted with $-i\boldsymbol\sigma$, where $\boldsymbol\sigma= (\sigma_x,\sigma_y,\sigma_z)$ is a vector of Pauli matrices. The quaternion wedge and dot products may be freely replaced by the vector cross and dot products respectively, and unit quaternions may be replaced by the corresponding unitary matrices, as seen below.  For a full review of all quaternion properties and operations used in this Paper, see App.~\ref{app:quat}. 

Before returning to the task of finding the fidelity and optimizing it, let us define the quantities $\Ei_i(t)$, given by  
\begin{eqnarray}
\Ei_i(t) &\equiv& -i\,\Ucal^{\dag}_c(t)\,\sigma_i \,\Ucal_c(t),
\end{eqnarray}
where, recalling that $\Ucal_c(t)$ is the unitary corresponding to the control Hamiltonian, $\Ei_i(t)$ are the Pauli matrices in the rotating frame of the control fields. Note that the $\Ei_i(t)$ satisfy the quaternion algebra, \emph{regardless\/} of the form of the control fields, for all times $t$, i.e.,
\begin{eqnarray}
\Ei_i(t)\Ei_j(t) &=& -\delta_{ij}\,\eins+\epsilon_{ij}^{\phantom{ij}k}\, \Ei_k(t), \nonumber\\
&& \label{eq:quat1}
\end{eqnarray}
where the indices $i$, $j$, and $k$ take on the values $1$, $2$, and $3$, corresponding respectively to $x$, $y$, and $z$, and repeated indices are summed over. The symbol $\eins$ denotes the $2\times2$ unit matrix.  Let us introduce $\Ei_0(t)\equiv\eins$; then evidently we have
\begin{equation}
\Ei_0(t)\Ei_i(t)=\Ei_i(t)\Ei_0(t) =\Ei_i(t) \label{eq:quat2}
\end{equation}
for all $i$.  The isomorphism between the set of quantities $\Ei_i(t)$ and quaternions is given by identifying $\Ei_i(t)$ (for $i=1,2,3$) respectively with the quaternion imaginary units $\hat{\imath},\hat{\jmath},\hat{k}$, and $\Ei_0(t)$ with the real number $1$.  Quaternion addition is defined component-wise, whereas multiplication is defined through Eqs.~(\ref{eq:quat1},\ref{eq:quat2}); note that quaternion multiplication is not commutative. 

Given this identification, for calculational purposes it is simpler to work with quaternion quantities.  Recall that any quaternion $Q$ can be expressed in terms of components as $Q=Q^i\,\Ei_i$ with $\left\{Q^i\right\}$ real and $i$ summed over the values $0,1,2,3$. The component $Q^0E_0$ is referred to as the \emph{real\/}, or \emph{scalar\/}, part of the quaternion, and the remainder is the \emph{imaginary\/} part, also called the \emph{vector} part, of the quaternion.  Quaternions having vanishing scalar part are called \emph{pure} quaternions. In what follows we will almost exclusively work with pure quaternions.
For additional properties of quaternions, other types of operations (specifically, the dot and wedge products) and the terminology associated with quaternions, we refer the reader to App.~\ref{app:quat}.  

We refer to the set of quaternions $\Ei_i(t)$, restricting $i$ to be $1,2,$ or $3$ as the \emph{rotating triad}, and we refer to the set $\ei_i$ given by
\begin{eqnarray}
\ei_i &=& -i\sigma_i, \label{eq:ei}
\end{eqnarray} 
as the \emph{static triad}, because these are fixed in the laboratory frame. %Most quantities we will be working with are \emph{pure} quaternions, i.e. they have vanishing scalar parts. 
 For convenience we represent \emph{pure quaternions\/} using a bold font, i.e., ${\bf p}$, and they can be represented by a restricted sum over the indices,  i.e., ${\bf p}=\sum_{i=1}^3p^i\, \Ei_i$ in the rotating basis and ${\bf p}=\sum_{i=1}^3p^i\,\ei_i$ in the static basis.  Unless indicated otherwise, from now on we interpret repeated indices as a summation over the set $i=1,2,3$.  

It is straightforward to find the relation between $\Ei_i(t)$ and $\ei_i$ within the quaternion language.  Associated with the evolution operator $\Ucal_c(t)$ is a \emph{unit} quaternion; for simplicity, we indicate unit quaternions by means of a regular font.  If we take the pure quaternion $\Om(t)$ to represent the control Hamiltonian in the quaternion language, $\Om(t)\equiv \omega^i(t)\,\Ei_i(t)$, then the associated unit quaternion, corresponding to the unitary in the vector language, is given by
\begin{eqnarray}
u_c(t) &=& \Tcal\mathrm{e}^{\frac{1}{2}\int_0^t\mathrm{d}t' \, \Om(t')}
\end{eqnarray}
and the relation between the rotating triad $\Ei_i(t)$ and the static triad $\ei_i$ is given by
\begin{eqnarray}
\Ei_i(t) &=& \bar{u}_c(t)\,\ei_i \, u_c(t) \label{eq:Ee}
\end{eqnarray}
where $\bar{u}_c(t)\equiv\Tcal\mathrm{e}^{-\frac{1}{2}\int_0^t\mathrm{d}t' \, \Om(t')}$ is the quaternion conjugate of $u_c(t)$ (corresponds to $\Ucal_c^{\dag}(t)$ in matrix language, see App.~\ref{app:quat}), 
and the operation on the right hand side is just quaternion multiplication.  In quaternion language, Eq.~(\ref{eq:Ee}) tells us that the $\ei_i$ and $\Ei_i(t)$ are related via a pure rotation --- in other words the rotating triad corresponds to a rigid rotation of the static triad.   

Let us now continue with the evaluation of the fidelity, but now making use of the quaternion language.  Equation~(\ref{eq:ahalf}) can be rewritten in terms of quaternions as 
\begin{eqnarray}
\Acal_{1/2}[{\bf E}(\cdot), n_i(\cdot)] &=& \mathrm{Sc}\,\Tcal\mathrm{e}^{\frac{\epsilon}{2}\int_0^{\tau}\mathrm{d}t \,{\bf n}(t) }, \label{eq:ahalf2}
\end{eqnarray}
where ${\bf n}(t)=n^i(t)\Ei_i(t)$ is now interpreted as the \emph{pure} quaternion representing the stochastic field in the rotating frame, $\Acal_{1/2}$ is now interpreted to be a functional of the rotating triad [we take ${\bf E}(t)$ as shorthand for the set of $\Ei_i(t)$] as well as the stochastic fields $\left\{ n_i(\cdot)\right\}$, and the $\mathrm{Sc}$ operation simply takes the \emph{scalar\/} part of the expression following it.  

In this formulation, the idea is to find the best rotating triad ${\bf E}(t)$, i.e., the one that \emph{maximizes} the fidelity.  It may seem that we have not gained much from this reformulation. However, working with ${\bf E}(t)$ is in fact a much simpler task than working with the control field $\Om(\cdot)$ directly, as $\Om(\cdot)$ is buried inside time-ordered exponentials [see Eqs.~(\ref{eq:ahalf},\ref{eq:ucalc})]. As a result,  in practice one usually resorts to approximate schemes.  
In contrast, the set ${\bf E}(t)$ appears at the same level as the stochastic fields, and thus can be treated exactly. Furthermore, once we have found the best triad history ${\bf E}(t)$, it is straightforward to recover the control fields $\Om(t)$:
\begin{eqnarray}
\Om(t) &=& \frac{1}{2}\big(\partial_t{\bf E}_k(t)\big){\bf E}_k(t)\nonumber\\
&=& \frac{1}{2}\epsilon^{ijk}{\bf E}_i(t) \big({\bf E}_j(t)\cdot\partial_t {\bf E}_k(t)  \big). \label{eq:Om}
\end{eqnarray}
These are the control fields in the \emph{rotating frame}. Ultimately, the objects we are interested in are the control field in the \emph{laboratory frame}, $\boldsymbol\omega(t)=\omega^i(t)\,{\bf e}_i$, which is trivially found via
\begin{eqnarray}
\boldsymbol\omega(t) &=& u_c(t)\,\Om(t)\,  \bar{u}_c(t) \nonumber\\
&=& \frac{1}{2}\epsilon^{ijk}u_c(t)\,{\bf E}_i(t)\,\bar{u}_c(t)\big({\bf E}_j(t)\cdot\partial_t {\bf E}_k(t)  \big) \nonumber\\
&=&  \frac{1}{2}\epsilon^{ijk} \,{\bf e}_i \big({\bf E}_j(t)\cdot\partial_t {\bf E}_k(t)  \big).
\end{eqnarray} 
In other words, $\omega^i(t)=\boldsymbol\omega(t)\cdot {\bf e}_i=\Om(t)\cdot {\bf E}_i(t)$.
The lab frame components of the control field can be recovered directly from the rotating frame triad ${\bf E}(t)$, without the need to take the intermediate step of computing either $\Om(t)$ or $u_c(t)$: 
\begin{eqnarray}
\omega^i(t) &=& \frac{1}{2}\epsilon^{ijk}\Ei_j(t)\cdot \partial_t\Ei_k(t) \label{eq:omegaEE}
\end{eqnarray} 
Here and elsewhere, the dot denotes the quaternion dot product (see App.~\ref{app:quat}), and repeated indices are summed over. Note that Eq.~(\ref{eq:omegaEE}) is an \emph{exact\/} relation between $\omb(t)$ and $\Ei(t)$. 
%The quantity $\epsilon_{ijk}\omega^k(t)$ has a natural interpretation as a \emph{connection form} for the frame field corresponding to the triad ${\bf E}_i(t)$ - we will see more hints of this later.     

By taking the triad $\Ei(t)$ as the relevant degrees of freedom [as opposed to the control $\Om(t)$], we no longer have to worry about the time-ordered exponential associated with $\Om(t)$, but there still remains an overall time-ordering operator in front of the whole expression; see Eq.~(\ref{eq:ahalf}). 
For a pure quaternion ${\bf n}(t)$, one can always find \emph{another} pure quaternion ${\bf m}_{\epsilon}(t)$ such that the relation
\begin{eqnarray}
\Tcal\mathrm{e}^{\frac{\epsilon}{2}\int_0^\tau\mathrm{d}t\,{\bf n}(t)} &=& \mathrm{e}^{\frac{\epsilon}{2}{\bf m}_{\epsilon}(\tau)} \label{eq:entemt}
\end{eqnarray} 
is satisfied, and we note that ${\bf m}_{\epsilon}(t)$ is a function of $\epsilon$ as well as time. The expression for ${\bf m}_{\epsilon}(t)$ can be found, as is commonly done in quantum mechanics from the Magnus expansion; i.e., as an expansion of $\log \Tcal \exp (\epsilon X)$ (for some quantum operator $X$, i.e. see Ref.~\cite{blanes2008}) for a small parameter $\epsilon$. By using the quaternion formulation, however, it is straightforward to determine how the two quantities ${\bf n}(t)$ and ${\bf m}_{\epsilon}(t)$ are related to each other \emph{exactly\/}; this relationship comes in the form of a differential equation, viz.,    
\begin{widetext}
\begin{eqnarray}
\frac{d{\bf m}_{\epsilon}(t)}{dt}&=& {\bf n}(t)-\frac{\epsilon}{2}{\bf m}_{\epsilon}(t)\wedge {\bf n}(t) +\left(1-\frac{\epsilon m_{\epsilon}(t)}{2}\cot\frac{\epsilon m_{\epsilon}(t)}{2}  \right)
\widehat{\bf m}_{\epsilon}(t)\wedge\big(\widehat{\bf m}_{\epsilon}(t)\wedge {\bf n}(t)\big) \label{eq:dmdt}
\end{eqnarray}
\end{widetext}
where $m_{\epsilon}(t)\equiv |{\bf m}_{\epsilon}(t)|$ is the quaternion modulus, and ${\bf\hat{m}}_{\epsilon}(t)={\bf m}_{\epsilon}(t)/|{\bf m}_{\epsilon}(t)|$ is a unit pure quaternion and, as such, it can be interpreted as the \emph{direction} of ${\bf m}_{\epsilon}(t)$ (see App.~\ref{app:ntom} for a derivation of this result). Note that in the case of pure quaternions the quaternion wedge product acts just like the vector cross product.  If Eq.~(\ref{eq:dmdt}) is solved perturbatively in $\epsilon$, one recovers the Magnus expansion (see App.~\ref{app:ntom} for details), but the real power of Eq.~(\ref{eq:dmdt}) is that it is an exact relation: the solution of this differential equation is equivalent to the exact summation over \emph{all} terms in the Magnus expansion.  

After this lengthy detour, let us return to the task of analyzing the fidelity.  Making use of the results in Eqs.~(\ref{eq:ahalf2},\ref{eq:entemt}), we find 
\begin{eqnarray}
\Acal_{1/2}[\Ei(\cdot), {\bf m}_{\epsilon}(\cdot)] &=& \cos\left(\frac{\epsilon m_{\epsilon}(\tau)}{2}  \right)
\end{eqnarray}
which, in view of its simplicity relative to the expression in terms of $\left\{n_i(\cdot)\right\}$, we can use to find the exact expression for the fidelity amplitude for general spin $s$:
\begin{eqnarray}
\Acal_s(\Ei(t),{\bf m}_{\epsilon}(t)) &=& \frac{1}{2s+1}\sum_{j=-s}^s\mathrm{e}^{-ij\epsilon m_{\epsilon}(\tau)}. \label{eq:acals}
\end{eqnarray}
This enables us to construct a simple expression for the fidelity $\Fcal_s=\langle \Acal_s \rangle$, using Eqs.~(\ref{eq:acals},\ref{eq:ninj}):
\begin{eqnarray}
\Fcal_s[\Ei(\cdot)] &=& \frac{1}{2s+1} \sum_{j=-s}^s \mathrm{e}^{-(j\epsilon)^2 S[\Ei(\cdot)]+O(\epsilon^3)} \label{eq:fcals}
\end{eqnarray}    
where $S[\Ei(\cdot)]$ is a functional of the rotating triad, i.e., 
\begin{eqnarray}
S[\Ei(\cdot)] &=& \frac{1}{2}\int_0^{\tau}\int_0^{\tau}\mathrm{d}t\, \mathrm{d}t'\,\Ncal_{ij}(t,t')\,\Ei_i(t)\cdot\Ei_i(t'). \label{eq:se}
\end{eqnarray}
In the expression (\ref{eq:fcals}) for $\Fcal_s$, we make the assumption that $\epsilon$ is small, and only consider the leading-order term in $\epsilon$ in the exponent. Higher order contributions can be easily computed, see App.~\ref{app:ntom} for details on how this is accomplished.  

It is clear from Eq.~(\ref{eq:fcals}) that in order to maximize the fidelity, we need only minimize the functional $S[\Ei(\cdot)]$ --- note that this condition is \emph{independent} of the spin number $s$.  In other words, the functional form of the rotating triad $\Ei_i(t)$ (and therefore the control field; see Eq.~(\ref{eq:omegaEE})\,) that maximizes the fidelity is the \emph{same} for all values of $s$.  The fidelity itself, however, \emph{does} depend on $s$.

%******************

\subsection{Extremals and constraints}

The task that remains is to find the rotating triad $\Ei_i(t)$ that minimizes the functional $S[\Ei(\cdot)]$; cf.~Eq.~[\ref{eq:se}].  We look for solutions in the form of extremals of the action, i.e., we seek $\Ei_i(t)$ such that the variation $\delta S[\Ei(\cdot)]$ vanishes.  Note that the variations in the triad $\delta \Ei_i(t)$ are not fully arbitrary.  They are \emph{constrained} due to the fact that the triad has to rotate as a rigid body.  Simply put, the variations are constrained to take the form
\begin{eqnarray}
\delta \Ei_i(t) &=& \delta{\bf A}(t)\wedge \Ei_i(t), \label{eq:deltaA}
\end{eqnarray}
where $\delta {\bf A}(t)$ is now \emph{any} arbitrary pure quaternion. 

In addition to the rigidity constraints just stated, there are also \emph{experimental} constraints present that one should account for, such as limitations on the frequency, amplitude, etc.~that the control fields can take. Even in the absence of experimental considerations, it is natural to impose such constraints, if we are to make fair comparisons between different realizations of the control fields. For instance, if there is no limit to the amplitude of the control field, one may simply pick a large enough amplitude such that one can achieve the target operation over a time-scale much smaller than any time-scale associated with the noise $\Ncal_{ij}(t,t')$.  The greater the separation of time scales, the smaller will be the effective action $S$, and thus the higher will be the fidelity $\Fcal$.  In this case, there is no sense in comparing strong fields with weak ones, as strong fields always win.  

%%%%%%%%%%%%%%%%%COME BACK HERE

In order to make a sensible comparison between different choices for the control fields, we have to put some constraint on the solution space in which we seek trajectories for the triad $\Ei(t)$.  As an example, from a purely physical standpoint, one may choose to compare trajectories for which the \emph{total energy output} associated with the control field is \emph{prescribed\/}.  This quantity is given by the time-integral of the \emph{square} of the control field, so one has the following constraint~\footnote{Strictly speaking, one should use the \emph{lab frame} quantity $|\boldsymbol\omega(t)|^2$, but in our system due to rotational invariance $|\Om(t)|^2=|\boldsymbol\omega(t)|^2$ \emph{for any and all} realizations of the control field.}     
\begin{equation}
E_{\rm out\/}\equiv \int_0^{\tau}\mathrm{d}t\, \frac{|\Om(t)|^2}{2} = \mathrm{const.}
\end{equation}
In what follows, we shall also prescribe the transit time $\tau$. As we are fixing the energy output, if we were to make $\tau$ too small, there would not be enough time to achieve a given target.  Effective values for $\tau$ should be bounded from below in an $E_{\rm out\/}$ dependent way to ensure that we have access to all desired targets.  We shall seek optimal controls in the space of fixed $\tau$ and $E_{\rm out\/}$.   

To determine the optimal controls in this constrained space we seek minima corresponding to the constrained functional $S_c$, determined via
\begin{eqnarray}
S_c &\equiv& \frac{1}{2}\int_0^{\tau}\mathrm{d}t\int_0^{\tau}\mathrm{d}t'\left\{\Ncal_{ij}(t,t')\,\Ei_i(t)\cdot \Ei_j(t')\right.\nonumber\\ 
&& \left.+\lambda \, \delta(t-t')|\Om(t)|^2  \right\}, \label{eq:SC}
\end{eqnarray}           
where $\lambda$ is the Lagrange multiplier associated with the energy output constraint, and the output power $|\Om(t)|^2$ is determined entirely in terms of the triad $\Ei(t)$, via
\begin{eqnarray}
|\Om(t)|^2 &=& \frac{1}{8}\delta^{ik}\delta^{j\ell}\Big(\Ei_i(t)\cdot\partial_t \Ei_j(t)- \Ei_j(t)\cdot \partial_t\Ei_i(t)\Big) \nonumber\\ 
&& \times \Big( \Ei_k(t)\cdot\partial_t \Ei_\ell(t)- \Ei_\ell(t)\cdot \partial_t\Ei_k(t)  \Big) \nonumber\\
&=& \frac{1}{2}\epsilon^{ijk}\, \Ei_i(t)\cdot\Big(\partial_t \Ei_j(t)\wedge \partial_t \Ei_k(t)  \Big). \label{eq:Omsq}
\end{eqnarray}
Thus we see that $S_c$ is a functional of the triad $\Ei(t)$ and its first time-derivatives only.  To obtain the first line of Eq.~(\ref{eq:Omsq}) we use the expression for $\Om(t)$ in Eq.~(\ref{eq:Om}); the second line requires a little algebra. In practice it is simpler to work with the expression as given in the second line, because it is of lower order in the triad and its derivatives. Note that $\lambda|\Om(t)|^2/2$ has a natural interpretation as the \emph{kinetic energy} associated with the triad $\Ei(t)$, with the Lagrange multiplier $\lambda$ playing the role of inertia. 

The constraint present in Eq.~(\ref{eq:SC}) is just one type of a constraint that we may impose on the system. We are free to impose other types.  For instance, we can replace the Lagrange multiplier in Eq.~(\ref{eq:SC}) by a matrix, and even make that time dependent: 
\begin{displaymath}
\frac{1}{2}\int_0^{\tau}\mathrm{d}t \, \boldsymbol\omega(t)\cdot \lambda(t) \cdot \boldsymbol\omega(t). 
\end{displaymath}
This may be useful, e.g., in the situation where the control fields $\boldsymbol\omega(t)$ are constrained to lie in the $xy$ plane. In this case, we simply let $\lambda_{zz}$ tend to infinity.  We are also free to impose constraints on the derivatives of $\boldsymbol\omega(t)$, i.e., to impose a frequency cutoff.  Indeed, we have a lot of freedom on the types of constraints we may impose. An advantage of this method (based on extremals) is the ease with which one can implement them. 

%Interestingly, the expression in the second line of Eq.~[\ref{eq:Omsq}] has a similar appearance to the so called Wess-Zumino term which appears in the action for a spin degree of freedom. The significance of the Wess-Zumino term is that in the quantum case its coefficient must be quantized for topological reasons; indeed its existence is crucial for the quantization of spin in the path integral language.  Our expression, however, does not lead to a Wess-Zumino term in the action $S_c$ and does not carry the same topological implications.  It is true, however, that $\lambda|\Om(t)|^2/2$ has a natural interpretation as the \emph{kinetic energy} associated with the triad $\Ei(t)$, with the Lagrange multiplier $\lambda$ playing the role of inertia. 

For illustration purposes, we work with with the case of scalar, time-independent $\lambda$ in Eq.~(\ref{eq:SC}), imposing constraints on the total energy output. From the action $S_c$, we can obtain the Euler-Lagrange equation, which will give us the condition that the extremals must satisfy. For our case, we are specifically interested in the \emph{minima\/}. %Recall that the triad is constrained to rotate as a \emph{rigid body} so the variations $\delta \Ei_i(t)$ take the form $\delta \Ei_i(t) = \delta {\bf A}(t)\wedge \Ei_i(t)$ (with $\delta {\bf A}(t)$ arbitrary) in order to satisfy the constraint.  
Setting $\delta S_c/\delta {\bf A}(t)=0$ (see Eq.~(\ref{eq:deltaA})\,) we get the Euler-Lagrange equation for the triad,
\begin{eqnarray}
\lambda\partial_t \Om(t)+\Ei_i(t)\wedge\left[\int_0^\tau \mathrm{d}t'\Ncal_{ij}(t,t') \Ei_j(t')\right] &=& 0, \label{eq:elEi}
\end{eqnarray}
where, in writing down Eq.~(\ref{eq:elEi}), we used the fact that
\begin{eqnarray}
\partial_t \Om(t) &=& \frac{1}{2}\epsilon^{ijk}\,\Ei_i(t)\,\big[\Ei_j(t)\cdot \partial_t^2\Ei_k(t) \big]. \label{eq:Omdot}
\end{eqnarray} 
The form of Eq.~(\ref{eq:elEi}) fits naturally with the fact that $\Om(t)$ is the angular velocity of the triad, and indicates that the interpretation of $\lambda|\Om(t)|^2/2$ as the kinetic energy, with $\lambda$ playing the role of inertia, is correct. 

Given the connection between $\Om(t)$ and the set $\Ei_i(t)$ (see Eq.~(\ref{eq:Om})\,), we also have
\begin{eqnarray}
\partial_t \Ei_i(t) &=& \Ei_i(t) \wedge \Om(t).
\end{eqnarray}
It is useful to define the \emph{dual triad} $\Di_i(t)$, via 
\begin{eqnarray}
\Di_i(t) &=& \int_0^{\tau}\mathrm{d}t'\Ncal_{ij}(t,t')\Ei_j(t'). \label{eq:dual}
\end{eqnarray}
The solution of the set of coupled equations
\begin{eqnarray}
\partial_t \Om(t) &=& -\frac{1}{\lambda} \Ei_i(t)\wedge\Di_i(t), \nonumber\\
\partial_t \Ei_i(t) &=& \Ei_i(t) \wedge \Om(t), \label{eq:eomcoupled}
\end{eqnarray}
along with the appropriate set of boundary conditions, determines the optimal controls. Recall that we have a target operation $U_T$, which is to be satisfied at the transit time $t=\tau$.  For our case, $U_T$ is simply a rotation operator, and can therefore always be written as
\begin{eqnarray}
U_T &=& \mathrm{e}^{-i\frac{1}{2}\theta_T {\bf \hat{r}}, \cdot \boldsymbol\sigma}
\end{eqnarray}
where $\theta_T$ is the angle and ${\bf \hat{r}}$ is the axis of rotation.  It is easy to translate this into quaternion language and find the corresponding unit quaternion $u_T$ that gives the target rotation
\begin{eqnarray}
u_T &=& \mathrm{e}^{\frac{1}{2}{\bf q}_T},
\end{eqnarray}
where ${\bf q}_T=q^i\ei_i$ is the pure quaternion corresponding to $-i\theta_T {\bf \hat{r}}\cdot\boldsymbol\sigma$,  as determined by Eq.~(\ref{eq:ei}).  The quaternion modulus $|{\bf q}_T|$ corresponds to the angle of rotation, and ${\bf\hat{q}}_T={\bf q}_T/|{\bf q}_T|$ corresponds to the axis of rotation.  
The boundary conditions for the equation of motion Eq.~(\ref{eq:eomcoupled}) are then entirely determined from the target $u_T$, via
\begin{eqnarray}
\Ei_i(0) &=& \ei_i, \nonumber\\
\Ei_i(\tau) &=& \bar{u}_T\ei_iu_T. \label{eq:bcEE}
\end{eqnarray}

%From the expression in Eq.~[\ref{eq:Omdot}] we see that   
%In obtaining this equation we made use of the fact that the triad rotates as a \emph{rigid body}, so the variations $\delta \Ei_i(t)$ are constrained to take the form $\delta \Ei_i(t) = \delta {\bf A}(t)\wedge \Ei_i(t)$ with $\delta {\bf A}(t)$ arbitrary. 
%Eq.~[\ref{eq:elEi}] is a second order differo-integral equation for the triad $\Ei_i$. 

%With some algebra, it is possible to isolate the acceleration term $\partial_t^2\Ei_k(t)$; in doing so we obtain a more convenient expression to work with
%\begin{eqnarray}
%\partial_t^2 \Ei_i(t) &=& \Om(t)\wedge(\Om(t)\wedge \Ei_i(t)) \nonumber\\
%&&-\frac{1}{\lambda}\Ei_i(t)\wedge \left[\Ei_j(t)\wedge\left[\int_0^\tau\mathrm{d}t'\,\Ncal_{jk}(t,t') \Ei_k(t')\right]\right] \nonumber\\
%&& \label{eq:eomEi}
%\end{eqnarray}
%where we recall that the control $\Om(t)$ (see Eq.~[\ref{eq:Om}]) is a function of the triad and its first order derivatives in time only.   
%Relevant solutions to our minimization problem correspond to solving the boundary value problem for Eq.~[\ref{eq:eomEi}] with $\Ei_i(0)=\ei_i$ and $\Ei_i(\tau)=\bar{u}_c\ei_iu_c$, where $u_c$ represents the predetermined target operation. 

The presence of boundary conditions specific to the target rotation is an inconvenience when investigating generic properties of extremals.  It is easy to take care of \emph{arbitrary} boundary conditions once and for all by transforming to a suitably \emph{rotating frame}, which is described by some angular drift velocity $\Om_0(t)$, generally an arbitrary function of time.  The rotating-frame triad $\widetilde{\Ei}_i(t)$ is then given by
\begin{eqnarray}
\widetilde{\Ei}_i(t) &=& u_0(t)\,\Ei_i(t) \,\bar{u}_0(t),
\end{eqnarray} 
where $u_0(t)\equiv\Tcal \exp\left[\frac{1}{2}\int_0^t\mathrm{d}t'\, \Om_0(t')\right]$. Likewise, for any rotating-frame quantity $\widetilde{\bf X}$ (e.g., the control fields $\widetilde{\Om}(t)$\,), we have $\widetilde{\bf X}=u_0(t)\,{\bf X}\,\bar{u}_0(t)$. We can also define the covariant derivative for the rotating frame $\widetilde{\partial}_t$, through the relation
\begin{eqnarray}
\widetilde{\partial}_t{\bf X}(t) &=& \partial_t{\bf X}(t)-\Om_0(t)\wedge {\bf X}(t);
\end{eqnarray}
likewise, for the noise kernel we have the covariant integral operator
\begin{eqnarray}
\int_0^\tau\mathrm{d}t' \, \widetilde{\Ncal}_{ij}(t,t') {\bf X}_j(t') &=& \int_0^\tau\mathrm{d}t'\Ncal_{ij}(t,t')\times \nonumber\\
&& u_0(t,t')\,{\bf X}_j(t')\,\bar{u}_0(t,t')
\end{eqnarray}
where $u_0(t,t')\equiv u_0(t)\,\bar{u}_0(t')$.  In terms of the quantities carrying tildes, the Euler-Lagrange equations for $S_c$ can be written in a \emph{generally covariant\/} form, valid for \emph{arbitrary\/} time-dependent rotating frame $\Om_0(t)$:
\begin{eqnarray}
\tpartial_t \tEi_i(t) &=& \tEi_i(t)\wedge \tOm(t), \nonumber\\
\tpartial_t\tOm(t) &=& -\frac{1}{\lambda}\tEi_i(t)\wedge \tDi_i(t). \label{eq:eqEOM}
\end{eqnarray}
A consequence of the fact that the form of $S_c$ expressed in terms of $\widetilde{\Ei}_i(t)$ is \emph{invariant\/} under arbitrary $\Om_0(t)$, is that this general covariance carries over to all physical equations. I.e., the relation between $\widetilde{\Om}(t)$ and $\widetilde{\Ei}_i(t)$ takes the expected form
\begin{eqnarray}
\widetilde{\Om}(t) &=& \frac{1}{2}\epsilon^{ijk}\,\widetilde{\Ei}_i(t)\,\left[\widetilde{\Ei}_j(t)\cdot \widetilde{\partial}_t\widetilde{\Ei}_k(t)   \right].
\end{eqnarray}

We can take advantage of the freedom we have in choosing $\Om_0(t)$ to force the boundary conditions to take on a simple form,
\begin{equation}
\widetilde{\Ei}_i(0)=\widetilde{\Ei}_i(\tau)=\ei_i, \label{eq:BCS}
\end{equation}
 the only requirement being that in order to satisfy the boundary condition in the laboratory frame, we simply need $\big(u_0(0), u_0(\tau)\,\big)=\big(1,u_T\,\big)$, with $u_T$ being the target operation.  The solutions $\widetilde{\Ei}_i(t)$ then satisfy periodic boundary conditions, i.e., each member of the triad forms a \emph{closed loop\/} on the unit sphere. The only other constraint is that the triad rotates as a rigid body.  The loops are not constrained in any other way, i.e., they are free to cross each other an arbitrary number of times.  

Without loss of generality, we choose $\Om_0(t)$ to be a constant, $\Om_D$, so that all nontrivial dynamical behavior is displayed by the triad $\widetilde{\Ei}_i(t)$. This particular rotating frame, with constant $\Om_D$, corresponds to eliminating the \emph{drift} term whose effect is to take the triad $\Ei_i(t)$ through a \emph{free geodesic path} on the unit sphere connecting the starting point at $t=0$ with the target at $t=\tau$.  This procedure simply accounts for this trivial part of the evolution, allowing us to focus on corrections due to fluctuations in the environment. For this reason, it is advantageous to solve the Euler-Lagrange equations in this rotating frame.  We take a further step, and work with the \emph{difference\/} $\delta\tOm(t)\equiv \tOm(t)-\Om_D$, and thus the set of equations
\begin{subequations}
\begin{eqnarray}
\partial_t \delta\tOm(t) &=& \Om_D\wedge \delta\tOm(t)-\frac{1}{\lambda} \tEi_i(t)\wedge \tDi_i(t), \label{eq:deltomande1}\\
\partial_t \tEi_i(t) &=& \tEi_i(t) \wedge \delta\tOm(t), \label{eq:deltomande2}
\end{eqnarray}
\end{subequations}
satisfying boundary conditions given in Eq.~(\ref{eq:BCS}). This set of variables holds the advantage that the trajectories $\delta\tOm(t), \tEi_i(t)$, which describe deviations from the average drift, are \emph{small} in the limit where the noise matrix $\Ncal_{ij}(t,t')$ is small, and vanish in the limit of vanishing noise, making them a natural set of variables to work with.   

Before proceeding with examples, we note that the trajectories as determined from Eqs.~(\ref{eq:deltomande1},\ref{eq:deltomande2}) can belong to \emph{distinct topological sectors}. This is easy to see in the case where there is no noise, i.e. when $\Om(t)=\Om_D$, since the set of drift vectors $\Om_D^{(n)}=(f+2\pi n)\, \hat{\bf r}$ (for constant $f$ and integer $n$) give rise to trajectories for $\Ei_i(t)$ that satisfy the same boundary conditions (i.e. correspond to the same target), but differ in the number of times the triad $\Ei_i(t)$ winds around the axis determined by $\hat{\bf r}$.  These \emph{winding numbers}, $n$, then describe different topological sectors. The same situation arises in the case of nonzero noise; here one also obtains sets of trajectories, differing in winding number, but satisfying the same boundary conditions.  When seeking solutions to the equations of motion, one can thus also specify the topological sector of interest, which can without loss of generality be accounted for by the drift term $\Om_D^{(n)}$ (where there is no risk of confusion we omit the label $n$ for brevity).   

Finally, we mention that the \emph{laboratory frame} components of the control field trajectories, the quantities we are ultimately interested in, can be determined entirely in terms of the rotating frame quantities $\omega_i(t)=\boldsymbol\omega(t)\cdot \ei_i=\Big(\Om_D+\delta\widetilde{\Om}(t)\Big)\cdot \widetilde{\Ei}_i(t)$.   

\subsection{Examples}

\begin{figure*}
	\includegraphics[scale=0.75, angle=270, trim={4cm 0 3cm  3.5cm},clip]{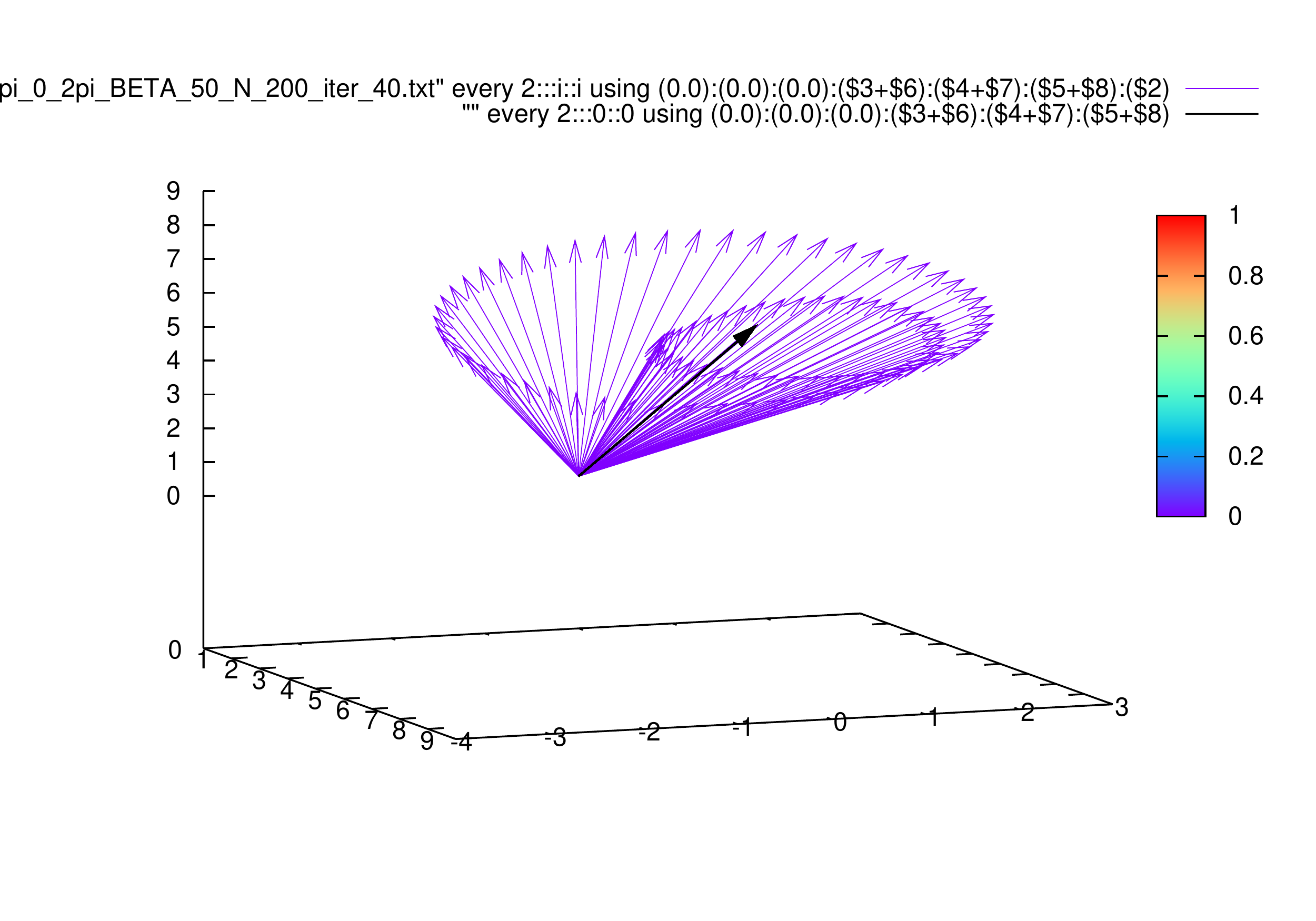}
		\caption{(Color online) Plot of $\boldsymbol\omega(t)$ (in units of $\tau^{-1}$) for $\lambda^{-1}=50\,\tau^{-1}$ (light purple arrows), for the case of $1/f$ noise.  The drift field $\Om_D=(2\pi,0,2\pi)/\tau$ (bold black arrow), which also corresponds to the optimal control for the case $\lambda^{-1}=0$, is shown for comparison. The other parameter values are $\gamma_1=20/\tau$, $\gamma_2=0.1/\tau$, and $\xi=8/\tau^2$, as described in the text.
 \label{fig:testomega} }
\end{figure*}

\begin{figure*}
	\includegraphics[scale=0.75]{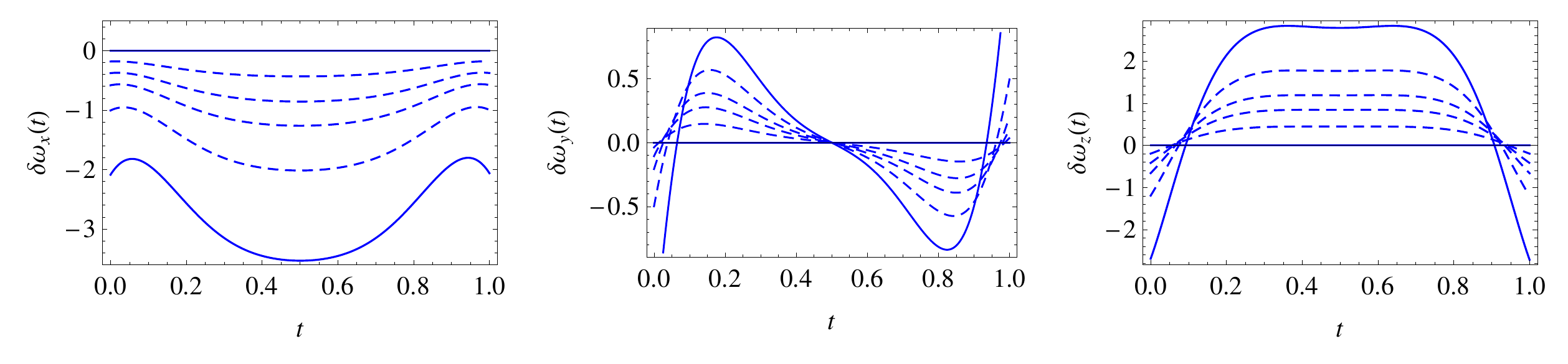}
		\caption{(Color online) Plot of the components of $\delta\boldsymbol\omega(t)$ (defined in text) vs.~time, for $\lambda^{-1}=0,10,20,30,50,100\,\tau^{-1}$.  The solid lines correspond to $\lambda^{-1}=100\tau^{-1}$ and the dashed  lines correspond to intermediate values. As explained in the text, note that the $x$ component of the control is reduced, relative to the noise-free case, at the expense of increasing the $z$ component. This reduces the component of the control field along the noise (taken to be the $x$ axis) in order to compensate more efficiently for the noise; see the text for further discussion.
 \label{fig:allomega} }
\end{figure*}

A physically interesting situation arises when the noise has a $1/f$ character. In many systems, this accounts for the material-dependent noise source that leads to decoherence in quantum devices~\cite{1freview}. It is therefore of considerate importance for applications to mitigate this noise effectively. $1/f$ noise can be understood as arising from the collective effect of multiple sources of \emph{telegraph noise\/} that are coupled to the quantum system of interest, with the weight of each source scaling as the inverse characteristic decay-rate of that source~\cite{1freview,bergli2009}.  

In what follows, we focus on the case of mitigating noise along a fixed axis, bearing in mind that generalizations are straightforward.  Without loss of generality, we take the noise to point along the $x$ axis, i.e., the only nonvanishing component of the noise matrix is $\Ncal_{xx}(t,t')$.  For $1/f$ noise, this is obtained via the expression
\begin{eqnarray}
\Ncal_{xx}(t,t') &=& \xi \int_{\gamma_1}^{\gamma_2} \frac{\mathrm{d}\gamma}{\gamma} \mathrm{e}^{-\gamma |t-t'|},
\end{eqnarray}
where $\xi$ denotes the strength of the noise, and $\gamma_1$ ($\gamma_2$) is the lower (upper) decay-rate cutoff for the ensemble of telegraph processes that are coupled to the spin. 
For frequencies $\nu$ within the range $ \gamma_1 \ll \nu \ll \gamma_2$, the noise correlator in the frequency domain $\Ncal_{xx}(\nu) \sim \nu^{-1}$, i.e. is $1/f$ in character.

For the noise form under consideration, we seek solutions for $\Om(t),\Ei_i(t)$ that dependend parametrically on the Lagrange multiplier $\lambda$.  The case $\lambda^{-1}=0$ corresponds to making energy output infinitely costly, and gives rise to solutions corresponding to the geodesic path on a sphere, i.e., we have $\Om(t)=\Om_D$.  For increasing $\lambda^{-1}$, the energy cost decreases, so that there is nontrivial competition between keeping energy costs low and compensating for the noise.  In the limiting case of $\lambda^{-1} \to \infty$, the energy costs become vanishingly small meaning that there are no constraints on the set of extremal control fields we get to choose from.  In this case, we can simply take the control $\Om(t)$ to be an infinitely sharp pulse acting over a vanishingly small transit time $\tau$, and this guarantees maximal fidelity. In actuality, as long as we ensure that the spectral content of the noise and the control do not overlap, which we are always free to do in the case where there are no constraints on the control, we guarantee maximal fidelity at leading order in $\epsilon$;  see Refs.~\cite{kofman2001,kofman2004}.  

As we have mentioned, one advantage of this approach is that we can naturally find families of solutions corresponding to a given $\Om_D$.  Raising $\lambda^{-1}=0$, we find a set of solutions that correspond to continuous deformations of  the geodesic on a sphere. As we raise $\lambda^{-1}$ and loosen the constraints on the energy output, we find solutions which improve the fidelity more and more.  
%There is no limit to how high we can raise the fidelity as long as we are willing to raise $\lambda^{-1}$, i.e. loosen the constraints of energy consumption.

In order to illustrate our approach, we take the drift $\Om_D$ to be $(2\pi,0,2\pi)/\tau$ so that we have a nontrivial winding-number (note that $\Om_D=(\sqrt{2}-1)(2\pi,0,2\pi)/\tau$ would give us the same target rotation), and the target operation has both a nonvanishing component \emph{parallel\/} to the noise and a nonvanishing component \emph{orthogonal\/} to it.  In this case, we expect to obtain nontrivial time dependence in both the amplitude of the control field, $|\Om(t)|$, and the direction $\widehat{\Om}(t)$.  
It is convenient to scale all parameters with respect to the transit time $\tau$, i.e., we take $\xi=8/\tau^2$ (the strength of the noise relative to the control is then $\sim \epsilon \xi$), $\gamma_1=0.1/\tau$, and $\gamma_2=20/\tau$. 
%since we are guaranteed to get data collapse by fixing the dimensionless combinations $\xi \tau^2$, $\gamma \tau$, and $\lambda/\tau$.  
The values chosen for the cutoffs $\gamma_1$ and $\gamma_2$ give us a wide range of frequencies for which the noise $\Ncal_{xx}$ is well approximated by $1/f$ noise.

Our interest lies in the case where $\lambda^{-1}$  is not too large, so that there is nontrivial competition between minimizing energy output and maximizing the fidelity.  The equations of motion (\ref{eq:eqEOM}) can be solved straightforwardly, numerically.  The quantities we are ultimately interested in are the \emph{control fields in the laboratory frame\/} $\boldsymbol\omega(t)$, whose components are given by 
\begin{eqnarray}
\omega_i(t) &=& \Om(t) \cdot \Ei_i(t)\nonumber\\
 &=& (\Om_D+\delta \tOm(t))\cdot \tEi_i(t), 
\end{eqnarray}
where $\Om_D=(2\pi,0,2\pi)/\tau$ is the drift and $\delta\tOm(t)\equiv \tOm(t)-\Om_D$, with $\tOm(t)$ and $\tEi_i(t)$ the control fields and the triad as given in the rotating frame; which we find by solving the rotating-frame Euler-Lagrange equations in the presence of the drift term $\Om_D$ (see Eqs.~(\ref{eq:deltomande1},\ref{eq:deltomande2}) and the accompanying discussion).
The optimal control field history $\boldsymbol\omega(t)$ in this nontrivial regime (with parameters given in the caption) is shown in Fig.~\ref{fig:testomega} (set of light purple arrows), where we have set $\lambda^{-1}=50/\tau$. The drift field $\Om_D$ ($(2\pi,0,2\pi)/\tau$) is shown (bold black arrow) for comparison.  

To get a better understanding of the results obtained in this example, let us take a closer look at how each component of $\boldsymbol\omega(t)$ behaves.  In Fig.~\ref{fig:allomega}, we plot the results for $\delta\boldsymbol\omega(t)\equiv \boldsymbol\omega(t)-\Om_D$.  The curves shown there correspond to several values of the Lagrange multiplier: $\lambda^{-1}=(0,10,20,30,50,100)/\tau$; in each plot the solid curve corresponds to $\lambda^{-1}=100/\tau$, and the dashed curves correspond to a sequence of smaller values. 

Recalling the fact that in this example the noise lies along the $x$ axis, the interpretation of the plots becomes clear: $\delta\omega_x(t)$ is \emph{negative definite\/}, meaning that the overall amplitude of the control field is \emph{reduced} relative to the drift $\Om_D$ for the \emph{entire process}.  This is reasonable because a driving field along the $x$ axis does nothing to compensate for noise along the $x$ direction: the driving and noise terms would commute in this case.  To compensate for the reduced drive along the $x$ axis, the drive along the $z$ axis on average is increased.  As far as the drive along the $y$ direction is concerned, although its time average is zero, it takes on a nontrivial time-dependence. 

 The behavior of the fields vary in concert in order to satisfy the boundary conditions whilst reducing the weight of the control field $\boldsymbol\omega(t)$ along the direction of the noise (in this case the $x$ direction) as much as possible.   As $\lambda^{-1}$ is increased, more and more of the weight lies along the $y$ and $z$ axes.  As $\lambda^{-1}$ is increased, and thus the energy restrictions are lessened, we have at our disposal larger-amplitude control fields $\boldsymbol\omega(t)$, and therefore larger frequencies in the spectral content of $\Ei_i(t)$.  In light of the fact that $1/f$ noise has higher weight at smaller frequencies,  by increasing $\lambda^{-1}$ we \emph{reduce} the spectral overlap between the noise and the control, and hence \emph{increase} the fidelity, which is what we set out to do.  To recapitulate, for the fidelity to be as large as possible, one should put as much of weight of the control field in a direction orthogonal to the noise for as long as possible.  This all has to be done in such a way as to satisfy the boundary conditions (i.e., to obtain the sought after target operation) and the energy constraints.  This is why we can only completely eliminate $\omega_x(t)$ in favor of $\omega_{y,z}(t)$ if we have access to arbitrarily large energy outputs. 

\begin{figure}
	\includegraphics[scale=0.65]{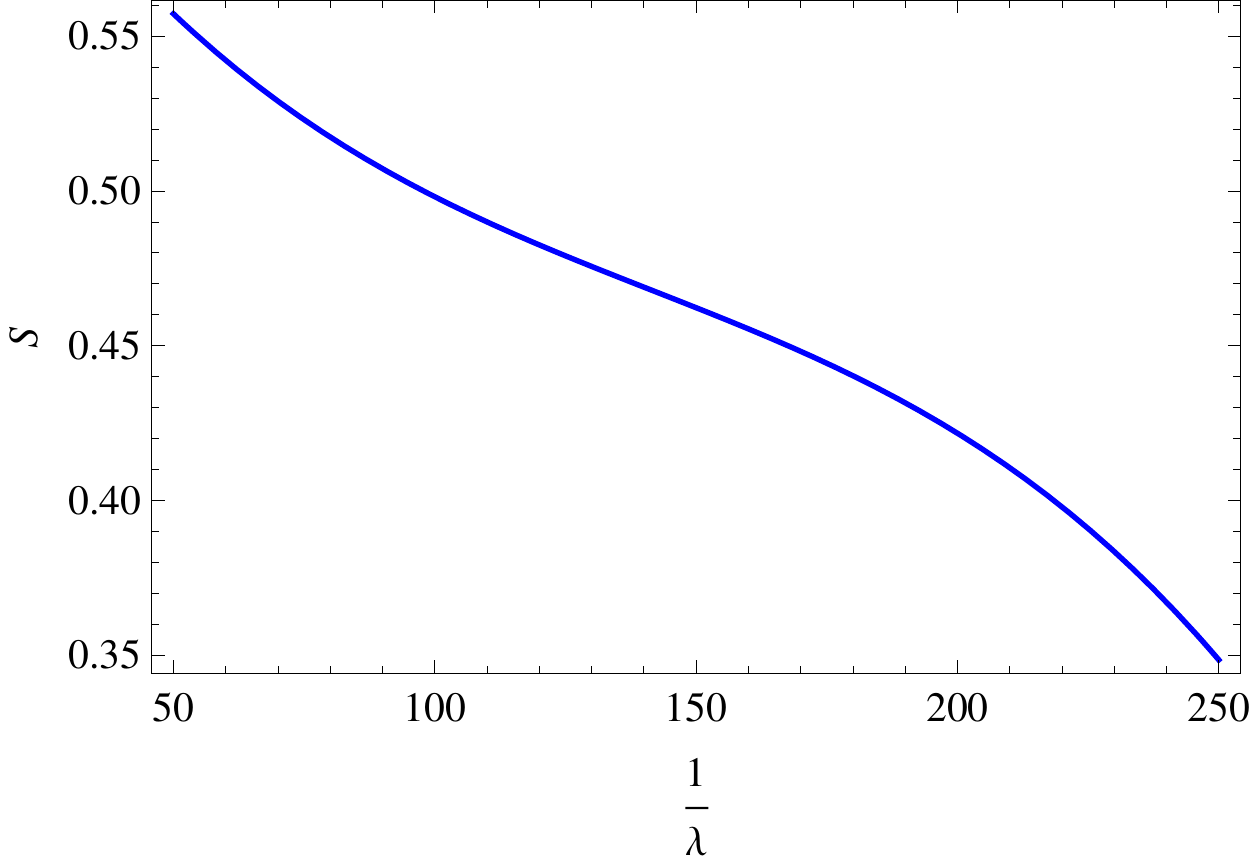}
		\caption{(Color online) Plot of the action $S$ as a function of $\lambda^{-1}$, which determines the constraint on energy output for the control fields $\Om(t)$.  Larger values of $\lambda^{-1}$ correspond to less stringent constraints, i.e., larger energy outputs. As $S$ approaches zero, the fidelity approaches unity (see Eq.~(\ref{eq:finalfcals})\,). There is no limit to how small we can make $S$ as long as we are willing to increase $\lambda^{-1}$. \label{fig:actionslamb} }
\end{figure}

Recall that although the optimal control scheme $\Big(\Om(\cdot),\Ei_i(\cdot)\Big)$ is independent of the spin quantum number $s$, the fidelity itself, however, is not, although the dependence on $s$ is elementary in the limit of weak noise:
\begin{equation}
\Fcal_s[\Ei(\cdot)] = \frac{1}{2s+1} \sum_{j=-s}^s \mathrm{e}^{-(j\epsilon)^2 S[\Ei(\cdot)]+O(\epsilon^3)}. \label{eq:finalfcals}
\end{equation}
Thus, in the small-$\epsilon$ limit considered here, one does best by reducing $S=\frac{1}{2}\int_0^{\tau}\mathrm{d}t\,\int_0^{\tau}\mathrm{d}t'\,\Ncal_{ij}(t,t')\Ei_i(t)\cdot \Ei_j(t')$ as much at the constraint in energy output will allow, obtaining an optimal control trajectory which is independent of $\epsilon$. For a fixed value  of $\epsilon$ (which we take to be $0.1$), we plot $S$ as a function of $\lambda^{-1}$ in Fig.~\ref{fig:actionslamb}.  Note that, in principle, there is no limit to how small $S$ can be as long as we are willling to keep increasing $\lambda^{-1}$.  For the case of $\lambda^{-1}=250/\tau$, the maximum value present in Fig.~\ref{fig:actionslamb}, the fidelity for the $s=1/2$ case is greater than $.999$. 

All results arrived at for this example can be generalized straightforwardly to the case of a general noise matrix. Greater fidelity usually results from putting as much of the weight of the control fields in a direction orthogonal to the \emph{dominant} noise contributions, while choosing the amplitudes $|\boldsymbol\omega(t)|$ such that the triad $\Ei_i(t)$ has as little spectral overlap with the noise as possible, in other words, as much as is allowed by the constraints imposed upon the control fields.  
%Our methods allow us to find the configurations for the control $\boldsymbol\omega(t)$ in a physically transparent, straightforward manner.  Another advantage afforded to us by this method is the fact that it is quite easy to incorporate a wide class of constraints into the problem.   

\section{Conclusions} \label{sec:conclusion}

We have considered the problem of error mitigation in the control of quantum systems subject to time-dependent sources of noise, which in general includes environmental noise and noise inherent to the experimental control apparatus. We do this in terms of a fidelity metric, which measures how faithfully the time evolution matrix (determined by the controls, internal system dynamics, and the noise) reproduces a predetermined `target' unitary transformation at the end of a prescribed transit time $\tau$.  We have tackled the problem through a modification of Schwinger-Keldysh path-integral techniques, which enables us to account for the effects of general sources of noise in a unified manner.  By analogy with the Martin-Siggia-Rose scheme we `integrate out' noise sources, thus arriving at an effective deterministic formulation, which we use to represent the fidelity. 
%which is essentially the overlap between the target operation we wish to impart upon the quantum system in question, and the \emph{actual} evolution acquired through the use of control fields in the presence of noise.  

Our methods yield the conditions obeyed by the control field history such that the effects of noise are optimally mitigated and hence, on average, the fidelity corresponding to the desired unitary transformation is as large as possible. 
%Our methods allow us to find the necessary conditions satisfied by the control fields in order to minimize the influence of the noise and at the same time achieve the target operation. Through the use of the path-integral language we derive an effective action principle where the control fields appear as effective degrees of freedom and the cumulants of the noise distribution appear as effective (time dependent) parameters in the action. 
These conditions are determined by solving equations of motion that are found by extremizing an effective action functional with the control fields playing the role of the degrees of freedom.  Our method has the advantage that it admits a wide range of constraints through the use of Lagrange multipliers.  These constraints may correspond to those naturally found in experiments, such as optimization of fidelity in the presence of fixed energy input, which is the main type of constraint we consider in the article.  

As an application of our methods, we consider a system composed of a single spin degree of freedom ${\bf \hat{S}}$ of arbitrary spin quantum number $s$ coupled to a noise source that is of the $1/f$ type, which is known to be a common source of quantum decoherence in many systems of interest.  We address the limit of weak noise, and study the problem in the case where the total energy output is constrained at a fixed value.  We use our methods to find the optimal control field histories subject to applied constraints, and interpret the results.  In the case of weak noise, we find that the optimal control field histories are independent of the spin quantum number $s$, and that the fidelity depends on $s$ in an elementary manner.    

Although we have studied the case of a single spin as a specific example, it is straightforward to generalize our methods, e.g., by applying them to chains of coupled spins, atomic systems, as well as general noise distributions.  An intriguing avenue that merits exploration, and that we have not pursued in this Paper, is the question of many body effects on the control task, where interactions may have nontrivial repercussions. This is a question that can be addressed within the formalism we have constructed here.

% One intriguing possibility we wish to explore further is the fact that through judicious choice of controls it may be possible to exploit many body effects as an aid to \emph{enhance} the fidelity (or other measures of interest such as entanglement which may be relevant in applications concerning quantum teleportation), a possibility we are actively exploring.  

\begin{acknowledgments}
We thank Kenneth R. Brown, Chingiz Kabytayev and J. True Merrill for valuable discussions. This work was supported by NSF grant DMR 12-07026 and by the Georgia Institute of Technology. Part of it was performed at the Aspen Center for Physics, which is supported by NSF grant PHY 10-66293.
\end{acknowledgments}

\appendix

\section{Quaternion algebra} \label{app:quat}
Here we briefly review quaternion algebra and explain the notation used in our Paper.
A quaternion $q$ can be written in terms of its components as follows (using the summation convention over repeated indices $\mu$)
\begin{eqnarray}
\mathfrak{q} &=&  q^{\mu} \ei_{\mu}\nonumber\\
&=& q^0\ei_0+q^1\ei_1+q^2\ei_2+q^3\ei_3,  \label{eq:appquat}
\end{eqnarray}
where the \emph{components} $q^{\mu}$ are real numbers, and the tetrad $\ei_{\mu}$ are the \emph{basis quaternions}.  The tetrad $\ei_{\mu}$ satisfy the quaternion multiplication table ($i$, $j$, and $k$ are restricted to take nonzero values):
\begin{eqnarray}
\ei_0\ei_i &=& \ei_i\ei_0=\ei_i \nonumber\\
\ei_i\ei_j &=& -\delta_{ij} + \epsilon_{ijk}\ei_k. \label{eq:appalg}
\end{eqnarray}
We can without loss of generality take $\ei_0=1$; the quaternion algebra can then be understood entirely in terms of the equation on the second line of Eq.~(\ref{eq:appalg}), i.e. we only need to consider the \emph{triad\/} $\ei_i$, where $i=1,2,3$.  We refer to the zeroth component $q^0$ as the \emph{scalar part\/} (in the literature, it is often referred to as the \emph{real part\/} in analogy with complex numbers), and we refer to ${\bf q}\equiv q^i \ei_i $ as the \emph{vector part\/} (it is also, in analogy with complex numbers, the \emph{imaginary part\/}). In terms of its scalar and vector parts, we can write an arbitrary quaternion $q$ as
\begin{equation}
q=q^0+{\bf q}.
\end{equation}  
Taking the analogy with complex numbers another step, we also have the \emph{quaternion conjugate\/} $\bar{q}$, where the conjugation operation is defined through its action on the triad $\ei_i$.  We have 
\begin{eqnarray}
\bar{\ei}_i &=& -\ei_i,
\end{eqnarray}
and in terms of an arbitrary quaternion written in terms of its scalar and vector parts, we then have
\begin{eqnarray}
\bar{q} &=& q^0 - {\bf q}.
\end{eqnarray}

In addition to the {\emph geometric product\/} between two quaternions $pq=r$, we can also define the quaternion dot product
\begin{eqnarray}
p\cdot q &\equiv& \frac{1}{2}(p\bar{q} + q\bar{p}),
\end{eqnarray}
and the quaternion wedge product
\begin{eqnarray}
p\wedge q &\equiv& \frac{1}{2} (pq-qp).
\end{eqnarray}
The \emph{magnitude\/} of a quaternion is given by its \emph{modulus\/} $|q|$, defined through the relation
\begin{equation}
|q| = \sqrt{q\bar{q}} = \sqrt{(q^0)^2+{\bf q}\cdot{\bf q}}.
\end{equation} 

Pure imaginary quaternions $\bf{q}$, referred to as \emph{pure quaternions}, can naturally be interpreted as a geometric vector in $\mathbb{R}^3$.  For pure quaternions, the quaternion dot and wedge products correspond to the usual vector dot and cross products respectively,
\begin{eqnarray}
{\bf p}\cdot {\bf q} &=& \frac{1}{2}p^iq^j(\ei_i \bar{\ei}_j+\ei_j\bar{\ei}_i) \nonumber\\
&=& p^iq^j \delta_{ij} \nonumber\\
{\bf p}\wedge {\bf q} &=& \frac{1}{2}p^i q^j (\ei_i\ei_j-\ei_j\ei_j) \nonumber\\
&=& p^iq^j \epsilon_{ij}^{\phantom{ij}k}\ei_k, 
\end{eqnarray}
where we use the quaternion multiplication table in Eq.~(\ref{eq:appalg}) (repeated indices are to be summed over).
The geometric quaternion product between two pure quaternions ${\bf pq}$ subsumes both the dot and wedge products
\begin{eqnarray}
{\bf pq} &=& -{\bf p}\cdot {\bf q} + {\bf p} \wedge {\bf q}.
\end{eqnarray}

\emph{Unit quaternions}, defined as quaternions for which $|u|=1$, can always be written in the form
\begin{eqnarray}
u&=& \mathrm{e}^{{\bf p}},
\end{eqnarray}
where ${\bf p}$ is a pure quaternion.  Unit quaternions are used in the quaternion language to describe rotations. Given a unit quaternion $v=\mathrm{e}^{\frac{1}{2}\boldsymbol\theta}$ and any pure quaternion ${\bf p}$ we can write down the rotated quaternion ${\bf p}'$ as
\begin{eqnarray}
{\bf p}' &=& v{\bf p}\bar{v} \nonumber\\
&=& \mathrm{e}^{\frac{1}{2}\boldsymbol\theta}{\bf p}\mathrm{e}^{-\frac{1}{2}\boldsymbol\theta} \nonumber\\
&=& {\bf p}\,\cos\theta + \widehat{\boldsymbol\theta}\wedge {\bf p}\, \sin\theta \nonumber\\
&&+ \big({\bf p}-(\widehat{\boldsymbol\theta}\cdot{\bf p})\widehat{\boldsymbol\theta}\big)\, (1-\cos \theta), \label{eq:appvpv}
\end{eqnarray}
where ${\bf p}'$, ${\bf p}$, and $\boldsymbol\theta$ are all pure quaternions, and where we have $\theta\equiv|\boldsymbol\theta|$ and $\widehat{\boldsymbol\theta}\equiv \boldsymbol\theta/|\boldsymbol\theta|$.  The third line of Eq.~(\ref{eq:appvpv}) is arrived at by the rules of quaternion algebra. 

The geometric interpretation of this relation is quite intuitive: $\boldsymbol\theta$ describes both the angle of rotation (through its modulus $\theta$), and the axis of rotation (through $\widehat{\boldsymbol\theta}$). Note that the quaternions $\pm v$ corresponds to the same rotation - this is analogous to the $2$ to $1$ correspondence between $SU(2)$ and $SO(3)$.  Indeed, there is an isomorphism between the description of rotations via unit quaternions and via $SU(2)$. An advantage of the quaternion description of rotations is that it does not run into issues of `gimbal lock' that afflict more conventional descriptions, such as Euler angles.  This robustness makes the quaternion description very attractive not just in a theoretical, but a practical point of view as well.

Finally, we note that composite rotations are also conveniently represented in terms of quaternions.  It is straightforward to show that the quaternion product of two unit quaternions is also a unit quaternion, and so can also be used to represent a rotation.  The quaternion corresponding to a rotation $u_1$ followed by a rotation $u_2$ is simply the quaternion product $u_3=u_2u_1$, which can be easily seen from the following
\begin{eqnarray}
{\bf p}'&=& u_3{\bf p}\bar{u}_3 \nonumber\\
&=& u_2u_1{\bf p}\overline{u_2u_1}\nonumber\\
&=& u_2(u_1{\bf p}\bar{u}_1)\bar{u}_2,
\end{eqnarray} 
where in going from the second to third lines, we used the following relation for the conjugate of the product of two quaternions $\overline{uv}=\bar{v}\bar{u}$.   

%**********************************************************************************

\section{Relation between $\nb(t)$ and $\mb_{\epsilon}(t)$ for $\Tcal\mathrm{e}^{\frac{\epsilon}{2}\int_0^t\mathrm{d}t'\, \nb(t')}=\mathrm{e}^{\frac{\epsilon}{2}\mb_{\epsilon}(t)}$}\label{app:ntom}

Our goal is to find the relation between the pure quaternions $\nb(t)$ and $\mb_{\epsilon}(t)$ (they can equivalently be considered members of ${\bf\mathfrak{su}}(2)$ due to the isomorphism between them) - we put a subscript $\epsilon$ in $\mb_{\epsilon}(t)$ as a reminder that it is generally a function of $\epsilon$.  In order to find the relation between $\nb(t)$ and $\mb_{\epsilon}(t)$, we make use of the defining relation
\begin{eqnarray}
\Tcal\mathrm{e}^{\frac{\epsilon}{2}\int_0^t\mathrm{d}t'\, \nb(t')}&=&\mathrm{e}^{\frac{\epsilon}{2}\mb_{\epsilon}(t)},
\end{eqnarray}
and the time derivative
\begin{eqnarray}
\frac{d}{dt}\Tcal\mathrm{e}^{\frac{\epsilon}{2}\int_0^t\mathrm{d}t'\, \nb(t')} &=& \frac{\epsilon}{2}\nb(t)\Tcal\mathrm{e}^{\frac{\epsilon}{2}\int_0^t\mathrm{d}t'\, \nb(t')},
\end{eqnarray}
in order to write the following expression
\begin{eqnarray}
\frac{\epsilon}{2}\nb(t) &=& \left[\frac{d}{dt}\mathrm{e}^{\frac{\epsilon}{2}\mb_{\epsilon}(t)}  \right]\mathrm{e}^{-\frac{\epsilon}{2}\mb_{\epsilon}(t)}. \label{eq:appnb1}
\end{eqnarray}
To find the derivative of the exponential, we first rewrite it as an infinite product and get the result
\begin{widetext}
\begin{eqnarray}
\frac{d}{dt} \mathrm{e}^{\frac{\epsilon}{2} \mb_{\epsilon}(t)} &=&\frac{d}{dt} \lim_{\substack{{N \to \infty}\\{\delta x\to 0}\\{N\delta x \to 1}}}\prod_{i=0}^N \mathrm{e}^{\frac{\epsilon}{2}\delta x\, \mb_{\epsilon}(t)}\nonumber\\
&=&\frac{\epsilon}{2} \lim_{\substack{{N \to \infty}\\{\delta x\to 0}\\{N\delta x\to 1}}} \sum_{k=0}^N \delta x\left[\prod_{i=0}^{N-k}\mathrm{e}^{\frac{\epsilon}{2}\delta x\, \mb_{\epsilon}(t)}\right] \frac{d\mb_{\epsilon}(t)}{dt} \left[\prod_{j=k}^N\mathrm{e}^{\frac{\epsilon}{2}\delta x\, \mb_{\epsilon}(t)} \right] \nonumber\\
&=&\frac{\epsilon}{2}\int_0^1\mathrm{d}x \, \mathrm{e}^{\frac{\epsilon}{2} x\, \mb_{\epsilon}(t)} \frac{d\mb_{\epsilon}(t)}{dt}
\mathrm{e}^{\frac{\epsilon}{2} (1-x)\, \mb_{\epsilon}(t)}.
\end{eqnarray}
\end{widetext} 
Using this expression, we can rewrite Eq.~(\ref{eq:appnb1}) as
\begin{eqnarray}
\nb(t) &=& \int_0^1\mathrm{d}x \, \mathrm{e}^{\frac{\epsilon}{2} x\, \mb_{\epsilon}(t)} \frac{d\mb_{\epsilon}(t)}{dt}
\mathrm{e}^{-\frac{\epsilon}{2} x\, \mb_{\epsilon}(t)}. \label{eq:appn2}
\end{eqnarray}
The geometric meaning of the integrand is simple: it corresponds to a rotation of $d\mb_{\epsilon}(t)/dt$ by an angle $\epsilon x m_{\epsilon}(t)$ (where $m_{\epsilon}(t)\equiv |\mb_{\epsilon}(t)|$) about the unit axis $\widehat{\mb}_{\epsilon}(t)\equiv \mb_{\epsilon}(t)/m_{\epsilon}(t)$.  Using quaternion algebra (or equivalently the ${\bf \mathfrak{su}}(2)$ algebra), we find
\begin{widetext}
\begin{eqnarray}
\mathrm{e}^{\frac{\epsilon}{2} x\, \mb_{\epsilon}(t)} \frac{d\mb_{\epsilon}(t)}{dt}
\mathrm{e}^{-\frac{\epsilon}{2} x\, \mb_{\epsilon}(t)} &=& \frac{d\mb_{\epsilon}(t)}{dt} + \widehat{\mb}_{\epsilon}(t)\wedge \frac{d\mb_{\epsilon}(t)}{dt} \, \sin x \epsilon m_{\epsilon}(t) + \widehat{\mb}_{\epsilon}(t) \wedge\left(\widehat{\mb}_{\epsilon}(t)\wedge\frac{d\mb_{\epsilon}(t)}{dt}  \right) (1-\cos x \epsilon m_{\epsilon}(t)). \nonumber\\
&&
\end{eqnarray}
\end{widetext}
With this, it is now trivial to carry out the integral in $x$ in Eq.~(\ref{eq:appn2}).  In order to simplify the expression further, we make use of the relation
\begin{eqnarray}
\frac{d\mb_{\epsilon}(t)}{dt} &=&\dot{ \widehat{\mb}}_{\epsilon}(t) m_{\epsilon}(t) + \widehat{\mb}_{\epsilon}(t) \dot{m}_{\epsilon}(t), \label{eq:appdmdt}
\end{eqnarray}
where the overhead dot denotes a time derivative, along with the fact that $\widehat{\mb}_{\epsilon}(t)$ and  $\dot{\widehat{\mb}}_{\epsilon}(t)$ are orthogonal since $\widehat{\mb}_{\epsilon}(t)$ is a unit pure quaternion.  We obtain the expression
\begin{eqnarray}
\nb(t) &=& \dot{m}_{\epsilon}(t) \widehat{\mb}_{\epsilon}(t) + \dot{\widehat{\mb}}_{\epsilon}(t) \frac{\sin \epsilon m_{\epsilon}(t)}{\epsilon}\nonumber\\
&&+\widehat{\mb}_{\epsilon}(t)\wedge \dot{\widehat{\mb}}_{\epsilon}(t) \frac{1-\cos\epsilon m_{\epsilon}(t)}{\epsilon} \label{eq:appn3}
\end{eqnarray}
This equation is very useful, as it gives us an exact expression for $\nb(t)$ in terms of $\mb_{\epsilon}(t)$ and its time derivatives.  Fortunately it is possible to invert Eq.~(\ref{eq:appn3}) and unambiguously solve for both $\dot{m}_{\epsilon}(t)$ and $\dot{\widehat{\mb}}_{\epsilon}(t)$ in terms of $\nb(t)$, $ m_{\epsilon}(t)$, and $\widehat{\mb}_{\epsilon}(t)$, where we find after some algebra
\begin{eqnarray}
\dot{m}_{\epsilon}(t) &=& \nb(t) \cdot \widehat{\mb}_{\epsilon}(t) \nonumber\\
\dot{\widehat{\mb}}_{\epsilon}(t) &=& -\frac{\epsilon}{2}\widehat{\mb}_{\epsilon}(t)\wedge \nb(t) \nonumber\\
&&-\left(\frac{\epsilon}{2}\cot\frac{\epsilon m_{\epsilon}(t)}{2}\right) \widehat{\mb}_{\epsilon}(t) \wedge\left(\widehat{\mb}_{\epsilon}(t)\wedge \nb(t)\right). \nonumber\\
&& \label{eq:appmmdot}
\end{eqnarray}
Using Eq.~(\ref{eq:appdmdt}), we can combine both equations in Eq.~(\ref{eq:appmmdot}) to find a single differential equation unambiguously relating $\mb_{\epsilon}(t)$ and $\nb(t)$.  After a little algebra we find
\begin{widetext}
\begin{eqnarray}
\frac{d\mb_{\epsilon}(t)}{dt} &=& \nb(t)-\frac{\epsilon}{2}\mb_{\epsilon}(t)\wedge \nb(t) +\left(1-\frac{\epsilon m_{\epsilon}(t)}{2}\cot\frac{\epsilon m_{\epsilon}(t)}{2}  \right) \widehat{\mb}_{\epsilon}(t)\wedge\left(  \widehat{\mb}_{\epsilon}(t)\wedge \nb(t) \right). \label{eq:appdmdtfinal}
\end{eqnarray}
\end{widetext}
Note that Eq.~(\ref{eq:appdmdtfinal}) is an exact relation between $\mb_{\epsilon}(t)$ and $\nb(t)$.  Solving it is equivalent to summing up all terms in the Magnus expansion for the case where $\nb(t)$ is an arbitrary time dependent linear combination of elements of ${\bf \mathfrak{su}}(2)$, or equivalently, any pure quaternion since the two are isomorphic.  For cases where $\epsilon$ is large, Eq.~(\ref{eq:appdmdtfinal}) can be used to give us the exact solution --- it can also be used to generate a diagrammatic expansion in $\epsilon$.  By summing up certain classes of diagrams, we can obtain useful results even for moderately large $\epsilon$.

For small $\epsilon$, we can solve Eq.~(\ref{eq:appdmdtfinal}) perturbatively in a fairly straightforward to arbitrary order.  This is useful if one is interested in finding a series expansion for higher order contributions for the fidelity functional.  We seek a solution of the form
\begin{eqnarray}
\mb_{\epsilon}(t) &=& \sum_{j=0}^{\infty}\epsilon^j \mb^{(j)}(t), \label{eq:appmbpert}
\end{eqnarray} 
where $\mb^{(j)}(t)$ is the $j$th order term in the perturbation theory (we drop the subscript $\epsilon$ for the perturbation terms $\mb^{(j)}(t)$ since the $\epsilon$ dependence is accounted for in the prefactor $\epsilon^j$).  Let us rewrite Eq.~(\ref{eq:appdmdtfinal}) in a way that is more amenable to this perturbative treatment.  We make use of the following series expansion~\cite{modernanalysis}
%~\footnote{We use the modern convention for Bernoulli numbers $B_j$, which is different from the one used in 
%, $B^{*}_j$. The relation between them is $B_{2j}=(-1)^{j-1}B^{*}_j (for $j$ integer)}
\begin{eqnarray}
1-\frac{x}{2}\cot\frac{x}{2} &=& \sum_{j=1}^{\infty}\frac{(-1)^{j+1}B_{2j}x^{2j}}{(2j)!}, \label{eq:appseriescot}
\end{eqnarray}  
where $B_{j}$ are the first Bernoulli numbers. 
The first Bernoulli numbers can be obtained from the Bernoulli polynomials, defined through the generating function
\begin{eqnarray}
\frac{x\mathrm{e}^{tx}}{\mathrm{e}^{x}-1}&=& \sum_{j=0}^{\infty}B_j(t)\frac{x^j}{j!},
\end{eqnarray}
where we have $B_j\equiv B_j(0)$, and are given explicitly by
\begin{eqnarray}
B_j &=& \sum_{k=0}^{j}\sum_{\ell=0}^{k}\frac{(-1)^{\ell}k!\ell^j}{\ell!(k-\ell)!(k+1)}.
\end{eqnarray}
Using Eq.~(\ref{eq:appseriescot}), we then see that the third term in Eq.~(\ref{eq:appdmdtfinal}) only contains positive even powers of $\epsilon m_{\epsilon}(t)$, where in particular the leading order term is quadratic in $\epsilon$.  In a series expansion in $\epsilon$, in the right hand side of Eq.~(\ref{eq:appdmdtfinal}), we get terms of the form
\begin{displaymath}
(-1)^{j+1}m_{\epsilon}^{2j}\widehat{\mb}_{\epsilon}(t)\wedge\left(  \widehat{\mb}_{\epsilon}(t)\wedge \nb(t) \right).
\end{displaymath}
These terms can be rewritten entirely in terms of $\mb_{\epsilon}(t)$ \emph{only}, simplifying the perturbative expansion since we do not have to consider the amplitude $m_{\epsilon}(t)$ and the unit axis $\widehat{\mb}_{\epsilon}(t)$ separately.  To do this, we introduce the \emph{nested} wedge product, which can be defined recursively through the relation
\begin{eqnarray}
({\bf a}\wedge)^j {\bf b} &\equiv& {\bf a}\wedge\left[({\bf a}\wedge)^{j-1} {\bf b} \right],
\end{eqnarray}
where $j$ is an integer greater than or equal to zero, and where we use the convention $({\bf a}\wedge)^0 {\bf b}={\bf b}$. Using this definition for the nested wedge product, along with some quaternion algebra, we can show that
\begin{eqnarray}
(-1)^{j+1}m_{\epsilon}^{2j}\widehat{\mb}_{\epsilon}(t)\wedge\left(  \widehat{\mb}_{\epsilon}(t)\wedge \nb(t) \right) &=& (\mb_{\epsilon}(t)\wedge)^{2j}\nb(t). \nonumber\\
&& \label{eq:appnest}
\end{eqnarray} 
Next, we take advantage of the fact that the odd Bernoulli numbers, $B_{2j+1}$, vanish for all integers $j>0$ so that we can include the odd order nested products in the series expansion without changing its value. Using this fact along with the result $B_0=1$ and $B_1=-1/2$ and the nested wedge product defined in  Eq.~(\ref{eq:appnest}), we can rewrite Eq.~(\ref{eq:appdmdtfinal}) in a simple unified way that is completely amenable to the perturbative expansion as given in Eq.~(\ref{eq:appmbpert}):
\begin{eqnarray}
\frac{d\mb_{\epsilon}(t)}{dt} &=& \sum_{j=0}^{\infty}\frac{\epsilon^{j}B_{j}}{j!}(\mb_{\epsilon}(t)\wedge)^{j}\nb(t).  \label{eq:appdmdt2}
\end{eqnarray}
We now define the nested wedge product for $N$ generally distinguishable factors, which can be, for example, factors $\mb^{(j_1)}(t), \mb^{(j_2)}(t),\cdots, \mb^{(j_N)}(t)$, appearing in the perturbative expansion.  It is defined recursively as
\begin{eqnarray}
\mb^{\{j_1j_2\cdots j_N\}}(t)\wedge\nb(t) &\equiv& \mb^{(j_1)}(t) \wedge\left[\mb^{\{j_2\cdots j_N\}}(t) \wedge \nb(t)  \right]. \nonumber\\
&& \label{eq:appnest2}
\end{eqnarray}
For the special case where the number of factors $N$ vanishes, i.e. $\{j_1\cdots j_N\}=\{ \emptyset \}$, we define $\mb^{\{\emptyset\}}(t)\wedge\nb(t)\equiv \nb(t)$ --- with this convention all other cases are uniquely defined through Eq.~(\ref{eq:appnest2}).  

Using Eqs.~(\ref{eq:appmbpert},\ref{eq:appdmdt2},\ref{eq:appnest2}), it is straightforward to find the expression for arbitrary $\mb^{(j)}(t)$ in the perturbative expansion.
The zeroth order $j=0$ term is given by the expression
\begin{eqnarray}
\mb^{(0)}(t) &=& \int_0^t\mathrm{d}t'\, \nb(t'), \label{eq:appmb0}
\end{eqnarray}
while for all other values $j> 0$ we find the result
\begin{widetext}
\begin{eqnarray}
\mb^{(j)}(t) &=&  \sum_{k=1}^{j}\frac{B_k}{k!}\left\{\sum_{\sum_{\ell=1}^{k}i_{\ell}=j-k}\int_0^t\mathrm{d}t' \, \mb^{\{i_1i_2\cdots i_N\}}(t')\wedge \nb(t') \right\}, \label{eq:appmbj}
\end{eqnarray}
\end{widetext}
where the second sum over the set $i_1,i_2,\cdots,i_k$ in Eq.~(\ref{eq:appmbj}) satisfies the constraint that $\sum_{\ell=1}^{k}i_{\ell}=j-k$.  One can easily check that with this constraint, only $\mb^{(k)}(t)$ for $0\leq k\leq j-1$ contribute to the right hand side of Eq.~(\ref{eq:appmbj}), so that this equation gives us an explicit result for the $j$th order perturbation term entirely in terms of lower order perturbation terms $\mb^{(k\leq j-1)}(t)$ and $\nb(t)$.  Eqs.~(\ref{eq:appmb0},\ref{eq:appmbj}) reproduce what is expected from the Magnus expansion~\cite{blanes2008}, which is not surprising since the same approximation scheme has been used here.  It can be easily shown that our perturbative result, Eqs.~(\ref{eq:appmb0},\ref{eq:appmbj}), readily generalizes to \emph{all} Lie algebras, provided one replaces the nested wedge products with nested commutators (Lie brackets) - in particular the quaternion wedge product corresponds exactly to the commutator for ${\bf \mathfrak{su}}(2)$.  For the case of quaternions (also ${\bf \mathfrak{su}}(2)$ due to the isomorphism), we have used quaternion methods to derive an \emph{exact} differential equation Eq.~(\ref{eq:appdmdtfinal}) in a straightforward manner, which when solved is equivalent to summing up all terms in the Magnus expansion.  The result given in  Eqs.~(\ref{eq:appmb0},\ref{eq:appmbj}), along with Eq.~(\ref{eq:appmbpert}), gives us a perturbative solution to the differential equation in powers of $\epsilon$ --- this reproduces the Magnus expansion term by term.     

Using Eqs.~(\ref{eq:appmb0},\ref{eq:appmbj}) we find explicit expressions for the first few terms for in the perturbative expansion, given here
\begin{widetext}
\begin{eqnarray}
\mb^{(0)}(t) &=& \int_0^t\mathrm{d}t_1\, \nb(t_1) \nonumber\\
\mb^{(1)}(t) &=& \frac{1}{2}\int_0^t\mathrm{d}t_1\int_0^{t_1}\mathrm{d}t_2 \, \nb(t_1)\wedge \nb(t_2) \nonumber\\
\mb^{(2)}(t) &=& \frac{1}{6}\int_0^{t}\mathrm{d}t_1\int_0^{t_1}\mathrm{d}t_2\int_0^{t_2}\mathrm{d}t_3 \left\{ \nb(t_1)\wedge [\nb(t_2)\wedge \nb(t_3)] + \nb(t_3)\wedge[\nb(t_2)\wedge \nb(t_1)] \right\} \nonumber\\
&& \mathrm{etc...} 
\end{eqnarray}
\end{widetext}

The form of Eq.~(\ref{eq:appdmdtfinal}) also suggests a different way of approximating the solution which is better than the Magnus expansion in the sense that we include contributions from \emph{all} orders of $\epsilon$ at each iteration (with the exception of zeroth order which coincides with the expression obtained from the Magnus expansion).  Because of this, we then obtain a much better approximation which works well even in the case were $\epsilon$ is large.  We denote the $[j]$th order approximation with a superscript $[n]$ using brackets instead of parentheses to distinguish this from the Magnus expansion (note that we now include the subscript $\epsilon$ since each term now \emph{does} depend explicitly on $\epsilon$).  
For the leading order we have
\begin{eqnarray}
\mb_{\epsilon}^{[0]}(t) &=& \int_0^{t}\mathrm{d}t' \, \nb(t'), 
\end{eqnarray}
and all higher order approximations $j>0$ are given by
\begin{widetext}
\begin{eqnarray}
\mb_{\epsilon}^{[j]}(t) &=& \int_0^{t}\mathrm{d}t'\left\{ \nb(t') - \frac{\epsilon}{2}\mb_{\epsilon}^{[j-1]}(t')\wedge \nb(t')  + \left(1-\frac{\epsilon m_{\epsilon}^{[j-1]}(t')}{2}\cot \frac{\epsilon m_{\epsilon}^{[j-1]}(t')}{2}  \right) \widehat{\mb}_{\epsilon}^{[j-1]}(t')\wedge (\widehat{\mb}_{\epsilon}^{[j-1]}(t')\wedge \nb(t')) \right\}. \nonumber\\
&&
\end{eqnarray}
\end{widetext}
The approximate $n$th order expression is simply given by the $n$th iterate of this procedure
\begin{eqnarray}
\mb_{\epsilon}(t) &\simeq& \mb_{\epsilon}^{[n]}(t).
\end{eqnarray}
This procedure is equivalent to solving the exact equation Eq.~(\ref{eq:appdmdtfinal}) by iteration --- when it is carried out to infinite order, it gives us the \emph{exact} solution (as long as the solution obtained this way is \emph{unique}, then it is unquestionably the solution), i.e. we have
\begin{eqnarray}
\mb_{\epsilon}(t) &=& \lim_{j \to \infty} \mb_{\epsilon}^{[j]}(t).
\end{eqnarray}
In practice we find that even for fairly large values of $\epsilon$, this procedure converges very quickly to a unique answer, i.e. there is some finite order $N$ upon which $\mb_{\epsilon}^{[N]}(t)$ is practically indistinguishable from the exact solution.  For extremely large values of $\epsilon$ this procedure may not converge --- i.e. the iterates may jump back and forth between two or more different values --- in which case one must modify the approximation to obtain a unique answer.  One can always check whether this unique answer is the solution by plugging it back into Eq.~(\ref{eq:appdmdtfinal}).

%%%%%%%%%%%%%%%%%%%%%%%%%%%%%%%%

\section{Evaluation of the path integral expression for the fidelity amplitude}\label{app:pathint}
%\end{widetext}
%Before proceeding with the evaluation of the path integral, consider the following change of variables which turns out to be advantageous for our purposes
%We first start by showing how to obtain the expression for the generating functional $G_s[J]$ in Eq.~ \ref{eq:gjbeval} (used to find the fidelity amplitude), which is expressed as a path integral.  We rewrite it here for convenience
%\begin{eqnarray*}
%G_s[\Jb] &=& \oint\frac{\mathrm{d}z}{2\pi i}\frac{\mathrm{e}^{\int_0^{\tau}\int_0^{\tau}\mathrm{d}t\mathrm{d}t'\;\Jb^{\star}(t) \Gcal_c(z,t,t')\Jb(t') }}{(2s+1)(1-z)^2z^{2s+1}}, 
%\end{eqnarray*}
%where we recall that the contour integral over $z$ is taken over any closed contour enclosing the pole at $z=0$, but not the pole at $z=1$.  

Here we show how to evaluate the path integral expression for the fidelity amplitude (given in Eq.~(\ref{eq:acali}) in the main Paper).  We rewrite it here for convenience (in this Appendix, unlike the main Paper, we write explicitly the normalization prefactor, $(2s+1)^{-1}$, corresponding to the dimension of the Hilbert space)
\begin{eqnarray}
\Acal_s &=& \frac{1}{2s+1} \mathrm{Tr}\int \textrm{D}^2\boldsymbol\alpha_f\textrm{D}^2\boldsymbol\alpha_b\times|\boldsymbol\alpha_b(0)\rangle\langle\boldsymbol\alpha_f(0)| \nonumber\\
&& \times \mathrm{e}^{-(|\boldsymbol\alpha_b(\tau)|^2+|\boldsymbol\alpha_f(0)|^2)+\boldsymbol\alpha_b^{\star}(\tau)\cdot \boldsymbol\alpha_f(\tau)} \nonumber\\
&&\times \mathrm{e}^{i\int_0^{\tau}\mathrm{d}t [\boldsymbol\alpha_f^{\star}\cdot (i\partial_t-\mathcal{H}_{\epsilon}) \cdot \boldsymbol\alpha_f-\boldsymbol\alpha_b^{\star}\cdot(i\partial_t-\mathcal{H}_c) \cdot \boldsymbol\alpha_b]}, \label{eq:appacali}
\end{eqnarray}
where $|\boldsymbol\alpha_{f,b}(0)\rangle$ are two-component \emph{coherent states\/} corresponding to the two-mode operator ${\bf \hat{a}}^{\dag}\equiv (\hat{a}^{\dag}_1, \hat{a}^{\dag}_2)$. 
The trace is taken over the complete set of states $|\psi\rangle$ satisfying the constraint $ \langle \psi|\frac{1}{2}{\bf \hat{a}}^{\dag} {\bf \hat{a}}  |\psi \rangle = s $, fixing the quantum spin number $s$.  The integral measure $\mathrm{D}^2\boldsymbol\alpha_i$ for $i=f,b$ is given by the following expression
\begin{eqnarray}
\mathrm{D}^2 \boldsymbol\alpha_i &\equiv& \lim_{N \to \infty} \prod_{n=0}^N \mathrm{d}^2\boldsymbol\alpha_i(t_n), \label{eq:appmeas}
\end{eqnarray} 
where $t_n=n\tau/N$ and $\tau$ denotes the total transit time.  The local (in time) measure $\mathrm{d}^2\boldsymbol\alpha_i(t_n)$ is given by
\begin{eqnarray}
\mathrm{d}^2\boldsymbol\alpha_i(t_n) &\equiv& \mathrm{d}^2\alpha_i^{(1)}(t_n)\,\mathrm{d}^2\alpha_i^{(2)}(t_n),
\end{eqnarray}
where the superscripts $(1)$ and $(2)$ denote the components of the two-component field $\boldsymbol\alpha_i$, and we have
\begin{eqnarray}
\mathrm{d}^2 \alpha_i^{(j)}(t_n) &\equiv& \frac{\mathrm{d}\alpha_i^{(j)}(t_n)\,\mathrm{d}\alpha_i^{\star(j)}(t_n)}{2\pi i} \nonumber\\
&=& \frac{\mathrm{d}\Re \alpha_i^{(j)}(t_n)\,\mathrm{d}\Im\alpha_i^{(j)}(t_n)}{\pi}, \label{eq:appmeasloc}
\end{eqnarray} 
where $\Re \alpha_i^{(j)}(t_n)$ and $\Im \alpha_i^{(j)}(t_n)$ denote the real and imaginary parts of $\alpha_i^{(j)}(t_n)$ respectively.
The first and second lines of Eq.~(\ref{eq:appmeasloc}) are equivalent, and it is a matter of convenience as to which representation we use. 

To evaluate Eq.~(\ref{eq:appacali}), we introduce the \emph{generating functional\/} $G_s[\Jb]$
\begin{eqnarray}
G_s[\Jb] &=& \frac{1}{2s+1} \mathrm{Tr}\int \textrm{D}^2\boldsymbol\alpha_f\textrm{D}^2\boldsymbol\alpha_b\times|\boldsymbol\alpha_b(0)\rangle\langle\boldsymbol\alpha_f(0)| \nonumber\\
&& \times \mathrm{e}^{-(|\boldsymbol\alpha_b(\tau)|^2+|\boldsymbol\alpha_f(0)|^2)+\boldsymbol\alpha_b^{\star}(\tau)\cdot \boldsymbol\alpha_f(\tau)} \nonumber\\
&&\times \mathrm{e}^{i\int_0^{\tau}\mathrm{d}t [\boldsymbol\alpha_f^{\star}\cdot (i\partial_t-\mathcal{H}_c) \cdot \boldsymbol\alpha_f-\boldsymbol\alpha_b^{\star}\cdot(i\partial_t-\mathcal{H}_c) \cdot \boldsymbol\alpha_b]} \nonumber\\
&& \times \mathrm{e}^{\int_0^{\tau}\mathrm{d}t [\Jb^{\star}\cdot\boldsymbol\alpha_f + \boldsymbol\alpha_f^{\star}\cdot\Jb]}, \label{eq:appgjb}
\end{eqnarray}
where $\Jb$ and $\Jb^{\star}$ are two-component \emph{source fields\/} which couple only to the \emph{forward branch\/} of the path integral (due to the fact that the noise field only couples to this branch --- see discussion in main Paper).  We can evaluate $\Acal_s$ (which includes the noise contribution) via the following expression
\begin{eqnarray}
\Acal_s &=& \exp\left\{-i\epsilon\int_0^{\tau}\mathrm{d}t \; \frac{\delta}{\delta\Jb(t)}\mathcal{H}_n(t) \frac{\delta}{\delta\Jb^{\star}(t)}\right\} G_s[\Jb]\Bigg|_{\substack{\Jb(t)=0 \\ \Jb^{\star}(t)=0}}. \nonumber\\
&& \label{eq:appacalder}
\end{eqnarray}
Note that $G_s[\Jb]$ depends only on the noise-free term in the Hamiltonian (which shows up symmetrically in both forward and backward branches). For this reason, it is a simpler task to calculate $G_s[\Jb]$ first, and then use Eq.~(\ref{eq:appacalder}) to obtain $\Acal_s$, as opposed to calculating $\Acal_s$ directly.  Taking the generating functional route also holds the advantage that the physics is more transparent.

Our first goal is to evaluate $G_s[\Jb]$ exactly, which we show in the following subsection. Once we have this quantity we will use it to obtain an exact expression for the fidelity amplitude $\Acal_s$.   

\subsection{Evaluation of the generating functional $G_s[\Jb]$}
In what follows, we shall evaluate the path integral expression for $G_s[\Jb]$ (defined in Eq.~(\ref{eq:appgjb})) \emph{exactly}\/.
We start with a change of variables (a \emph{Keldysh rotation}) defined by
\begin{eqnarray}
\boldsymbol\psi(t) &=& \frac{\boldsymbol\alpha_f(t) + \boldsymbol\alpha_b(t)}{2} \nonumber\\
\boldsymbol\eta(t) &=& \boldsymbol\alpha_f(t)-\boldsymbol\alpha_b(t) \label{eq:cov}
\end{eqnarray}
It is easy to check that the Jacobian associated with this change of variables is exactly unity.
The description of the path integral in terms of these fields has a nice physical interpretation.
The symmetric combination $\boldsymbol\psi$ is usually referred to as the `classical' component in the literature (it is the only component that survives in the classical limit, giving rise to a unique classical trajectory corresponding to the saddle point of $G_s[\Jb]$).  The antisymmetric combination $\boldsymbol\eta$ is referred to as the `quantum' component since it accounts for deviations from the classical trajectory~\cite{kamenev2011}.  The evaluation of the path integral in terms of this set of variables simplifies greatly as we shall see below. %(this is an exact statement for noninteracting systems - this separation between classical and quantum components is only approximately true in the presence of interactions).   %In the semiclassical limit, one can treat $\boldsymbol\eta$ perturbatively.
%in the classical limit, while the antisymmetric combination $\boldsymbol\eta$ is referred to as the `quantum' part since it accounts for the effect of quantum fluctuations.
%To aid us in our evaluation of Eq~[\ref{eq:appacali}], we introduce external \emph{source fields} which couple to our quantum system as follows:  $\Lb(t)$ (referred to as the \emph{quantum source}) couples to quantum component $\etab(t)$, and $\Kb(t)$ (the \emph{classical source}) couples to the classical component $\psib(t)$. 

Let us rewrite the expression for $G_s[\Jb]$ in terms of this new set of variables.
After some algebra (and integration by parts in order to move time derivatives from $\etab$'s to $\psib$'s) we obtain the following expression
\begin{eqnarray}
G_s[\Jb] &=& \int \textrm{D}^2\psib \textrm{D}^2\etab\, W(\psib_0,\psib^{\star}_0,\etab_0,\etab^{\star}_0)\, \mathrm{e}^{-\frac{1}{2}|\etab_{\tau}|^2}\nonumber\\
&&\times \mathrm{e}^{\int_0^{\tau}\mathrm{d}t\left[i\etab^{\star} \cdot(i\partial_t \psib-\mathcal{H}_c \cdot \psib-i\Jb/2)+\psib^{\star}\cdot \Jb \right]} \nonumber\\
&&\times  \mathrm{e}^{\int_0^{\tau}\mathrm{d}t\left[(-i\partial_t \psib^{\star}- \psib^{\star} \cdot\mathcal{H}_c -i\Jb^{\star}/2) \cdot i\etab+\Jb^{\star}\cdot \psib \right]}, \label{eq:appgrot}
\end{eqnarray}
where we use the shorthand notation $\psib_0\equiv \psib(0)$, $\etab_0\equiv\etab(0)$, and $\etab_{\tau}\equiv \etab(\tau)$.
The function $W(\psib_0,\psib^{\star}_0,\etab_0,\etab^{\star}_0)$ in the integrand of Eq.~(\ref{eq:appgrot}) is given by the expression
\begin{eqnarray}
W(\psib_0,\psib^{\star}_0,\etab_0,\etab^{\star}_0)&=&  \mathrm{e}^{\frac{1}{2}(\psib_0^{\star}\cdot\etab_0-\etab^{\star}_0\cdot\psib_0)} \nonumber\\
&& \times \frac{1}{2s+1}\mathrm{Tr}\,|\psib_0-\frac{\eta_0}{2}\rangle \langle \psib_0+\frac{\etab_0}{2}|, \nonumber\\
&& \label{eq:appweight}
\end{eqnarray}
where the states $|\psib_0\pm \frac{\etab_0}{2} \rangle$ are coherent states, and we recall that the trace in Eq.~(\ref{eq:appweight}) is constrained to be over states for which the spin quantum number $s$ is \emph{fixed\/}, i.e. the constrained subspace contains the two-mode Fock states $|n_1,n_2 \rangle$ for which $n_1+n_2=2s$.

Before evaluating the path integral expression for $G_s[\Jb]$ in Eq.~(\ref{eq:appgrot}), we take a moment to write down explicitly the form taken by the integral measures in the new set of variables, $\psib$ and $\etab$. We have 
\begin{equation}
\mathrm{D}^2 \boldsymbol\psib \equiv \lim_{N \to \infty} \prod_{n=0}^N \mathrm{d}^2\psib(t_n),  \qquad
\mathrm{D}^2 \boldsymbol\etab \equiv \lim_{N \to \infty} \prod_{n=0}^N \mathrm{d}^2\etab(t_n), \label{eq:appmeas2}
\end{equation} 
with the local measures $\mathrm{d}^2\psib(t_n)$ and $\mathrm{d}^2\etab(t_n)$ taking the expected form
\begin{eqnarray}
\mathrm{d}^2\psib(t_n) &\equiv& \mathrm{d}^2\psi^{(1)}(t_n) \, \mathrm{d}^2 \psi^{(2)}(t_n) \nonumber\\
\mathrm{d}^2\etab(t_n) &\equiv& \mathrm{d}^2\eta^{(1)}(t_n) \, \mathrm{d}^2 \eta^{(2)}(t_n), 
\end{eqnarray}
where as usual, the superscripts denote the components of the fields. Importantly, the expression for the measures $\mathrm{d}^2\psi^{(j)}(t_n)$ and $\mathrm{d}^2\eta^{(j)}(t_n)$ slightly differ. Suppressing time arguments for brevity, we have
\begin{align}
\mathrm{d}^2\psi^{(j)} & \begin{aligned}
\equiv \frac{\mathrm{d}\psi^{(j)}\,\mathrm{d}\psi^{\star(j)}}{\pi i}&= \frac{2\, \mathrm{d}\Re\psi^{(j)}\,\mathrm{d}\Im\psi^{(j)}}{\pi} 
\end{aligned}
\nonumber\\
\mathrm{d}^2\eta^{(j)}& \begin{aligned}
\equiv \frac{\mathrm{d}\eta^{(j)}\,\mathrm{d}\eta^{\star(j)}}{4\pi i} &= \frac{\mathrm{d}\Re\eta^{(j)}\, \mathrm{d}\Im\eta^{(j)}}{2\pi},
\end{aligned}
\end{align}
where we see that the measure for $\psib$ is a factor of four larger than the measure for $\etab$. Note that the expressions in the second and third columns are entirely equivalent representations, and one can simply choose whichever representation is most convenient when performing calculations. 

Going back to the expression for $G_s[\Jb]$ in Eq.~(\ref{eq:appgrot}), we can immediately evaluate the integrals over $\etab_0,\etab^{\star}_0,\etab_{\tau}$ and $\etab^{\star}_{\tau}$. The only contribution to the integrand for the fields $\etab_0,\etab^{\star}_0$ that survives the time-continuum limit of the path integral comes from the term $W(\psib_0,\psib^{\star}_0,\etab_0,\etab^{\star}_0)$ given in Eq.~(\ref{eq:appweight}), giving us the result    
\begin{eqnarray}
\mathcal{W}(\psib_0,\psib^{\star}_0)&\equiv& \int \mathrm{d}^2\etab_0  \, W(\psib_0,\psib^{\star}_0,\etab_0,\etab^{\star}_0) \nonumber\\
&=& \frac{(-1)^{2s}}{2s+1}L_{2s}^{(1)}(4|\psib_0|^2)\mathrm{e}^{-2|\psib_0|^2}, \label{eq:appwigner}
\end{eqnarray}
where $L_{2s}^{(1)}(x)$ is an associated Laguerre polynomial, which we express here in terms of its Rodrigues representation
\begin{eqnarray}
L_n^{(k)}(x) &=&  \frac{\mathrm{e}^x x^{-k}}{n!}\frac{d^n}{dx^n}(\mathrm{e}^{-x}x^{k+n}) \nonumber\\
&=& \sum_{j=0}^n \frac{(k+n)!}{(n-j)!(k+j)!} \frac{(-x)^j}{j!}.
\end{eqnarray}
In the process of evaluating Eq.~(\ref{eq:appwigner}), we made use of the following expression giving the overlap between the two-component coherent states $|\psib_0 \pm \frac{1}{2}\etab_0 \rangle$ and the two-component Fock states $|n,2s-n \rangle$:
\begin{eqnarray}
\langle n,2s-n| \psib_0 \pm \frac{1}{2}\etab_0 \rangle &=& \frac{\mathrm{e}^{-\frac{1}{2} |\psib_0\pm\frac{1}{2}\etab_0|^2}}{\sqrt{n!(2s-n)!}}\left(\psi_0^{(1)} \pm \frac{1}{2}\eta_0^{(1)} \right)^n \nonumber\\
&&\times \left(\psi_0^{(2)} \pm \frac{1}{2}\eta_0^{(2)}  \right)^{2s-n}, \label{eq:appoverlap}
\end{eqnarray}
where the superscripts $(1)$ and $(2)$ on the right hand side of Eq.~(\ref{eq:appoverlap}) denote the components of the respective fields.
The function $\mathcal{W}(\psib_0,\psib^{\star}_0)$ given in Eq.~(\ref{eq:appwigner}), referred to as the \emph{Wigner function\/} in the context of quantum dynamics, can be interpreted as a distribution over the fields $\psib_0,\psib^{\star}_0$. Note that this distribution function, as described by $\mathcal{W}(\psib_0,\psib^{\star}_0)$, is generally \emph{not\/} positive definite, and in our case it shows oscillatory behavior.  As far as the integral over $\etab_{\tau}, \etab_{\tau}^{\star}$ is concerned, the only contribution that survives the time-continuum limit is given by
\begin{eqnarray}
\int \mathrm{d}^2\etab_{\tau}\, \mathrm{e}^{-\frac{1}{2}|\etab_{\tau}|^2} &=& 1. \label{eq:apptriv}
\end{eqnarray}     

We now proceed to evaluate the integrals over the fields $\etab(t), \etab^{\star}(t)$ in the bulk (i.e. for all $t>0$ in the path integral given in Eq.~(\ref{eq:appgrot})). The change of variables introduced in Eq.~(\ref{eq:appgrot}) really simplifies things here.  The integrals are readily evaluated and result in a product of \emph{Dirac delta functions\/} 
\begin{gather}
\lim_{N \to \infty} \prod_{n=0}^{N-1} \Bigg[\delta\Big(\psib(t_{n+1})+i\delta t[\Hcal_c(t_n)\cdot \psib(t_n)+i\Jb(t_n)/2]\Big)  \nonumber\\
\times  \delta\Big(\psib^{\star}(t_{n+1})-i\delta t[\psib^{\star}(t_n)\cdot\Hcal_c(t_n)+i\Jb^{\star}(t_n)/2]\Big)(2\pi)^2 \Bigg], \label{eq:appdeltas}
\end{gather}
where $\delta t\equiv \tau/N$.
This result makes the evaluation of the integrals over $\psib(t)$ and $\psib^{\star}(t)$ (for $t>0$) simple as well.  The delta functions constrain the fields $\psib(t),\psib^{\dag}(t)$ (for $t>0$) to obey the following equations of motion
\begin{eqnarray}
i\partial_t \psib &=& \Hcal_c \cdot \psib+\frac{i\Jb}{2} \nonumber\\
i\partial_t \psib^{\star} &=& -\psib^{\star}\cdot \Hcal_c - \frac{i\Jb^{\star}}{2},
\end{eqnarray}
where we have suppressed time arguments for brevity.
Assuming general initial conditions $\psib(0)=\psib_0$ we can write down the exact solution for all $t>0$
\begin{eqnarray}
\psib(t) &=& \mathcal{U}_c(t,0) \cdot \psib_0 + \frac{1}{2}\int_0^{t}\mathrm{d}t' \, \mathcal{U}_c(t,t') \cdot \Jb(t') \nonumber\\
\psib^{\star}(t) &=& \psib^{\star}_0 \cdot \Ucal_c(0,t)-\frac{1}{2}\int_0^{t}\mathrm{d}t'\,\Jb^{\star}(t')\cdot \mathcal{U}_c(t',t), \nonumber\\
&&  \label{eq:appclass}
\end{eqnarray}
where we have
\begin{eqnarray}
\mathcal{U}_c(t,t') &=& \mathcal{T}\exp\left[-i \int_{t'}^t \mathcal{H}_c(t'')\mathrm{d}t''  \right].
\end{eqnarray}
Note that in general we have $\psib^{\star}(t) \neq [\psib(t)]^{\star}$, as one can see from Eq.~(\ref{eq:appclass}), and these fields have to be treated as being independent of one another.  This also applies to the fields $\psib^{\star}_0$ and $\psib_0$. 

Applying the results obtained in Eqs.~(\ref{eq:appwigner}), (\ref{eq:apptriv}), (\ref{eq:appdeltas}), and (\ref{eq:appclass}) to Eq.~(\ref{eq:appgrot}), we get the following expression for $G_s[\Jb]$:
\begin{widetext}
\begin{eqnarray}
G_s[\Jb] &=&\mathrm{e}^{\frac{1}{2}\int_0^{\tau}\mathrm{d}t\int_0^{t}\mathrm{d}t' \, [\Jb^{\star}(t)\cdot \Ucal_c(t,t')\cdot \Jb(t') - \Jb^{\star}(t')\cdot\Ucal_c(t',t)\cdot \Jb(t)]} \times \int\mathrm{d}^2\psib_0 \, \mathcal{W}(\psib_0,\psib^{\star}_0) \,\mathrm{e}^{\int_0^{\tau}\mathrm{d}t\, [\Jb^{\star}(t)\cdot\Ucal_c(t,0)\cdot \psib_0+\psib_0^{\star}\cdot \Ucal_c(0,t)\cdot \Jb(t)]}. \nonumber\\
&& \label{eq:applong}
\end{eqnarray}
\end{widetext}
All that remains to be done in order to obtain the final result for $G_s[\Jb]$ is the evaluation of ordinary integrals over the fields $\psib_0$ and $\psib_0^{\star}$. This is a Gaussian integral with a polynomial prefactor in the integrand (see Eq.~(\ref{eq:appwigner})), and it can therefore  be evaluated exactly. Before doing so we rewrite the expression for $\mathcal{W}(\psib_0,\psib_0^{\star})$ by making use of the following integral representation for the associated Laguerre polynomial
 \begin{eqnarray}
L_n^{(k)}(x) = \oint \frac{\mathrm{d}z}{2\pi i} \, \frac{(-1)^{k}\mathrm{e}^{\frac{xz}{1+z}}}{(1+z)^{k+1}z^{n+1}}, \label{eq:applagint}
\end{eqnarray}  
where the contour runs counterclockwise and encloses the pole at $z=0$, but not the pole at $z=1$.  We can use the expression in Eq.~(\ref{eq:applagint}) to rewrite $\mathcal{W}(\psib_0,\psib_0^{\star})$ as follows
\begin{eqnarray}
\mathcal{W}(\psib_0,\psib_0^{\star}) &=& \frac{1}{2s+1}\oint\frac{\mathrm{d}z}{2\pi i}\frac{\mathrm{e}^{-2\left(\frac{1-z}{1+z} \right)|\psib_0|^2}}{(1+z)^2 z^{2s+1}}
\end{eqnarray}
This greatly simplifies the evaluation of the integrals over $\psib_0$ and $\psib^{\star}_0$ in Eq.~(\ref{eq:applong}), and we obtain the following result:
\begin{eqnarray}
\int \mathrm{d}^2\psib_0 \cdots &=& \oint\frac{\mathrm{d}z}{2\pi i}\frac{\mathrm{e}^{\frac{1}{2}\left(\frac{1+z}{1-z}\right)\int_0^{\tau}\int_0^{\tau}\mathrm{d}t\,\mathrm{d}t' \,\Jb^{\star}(t)\cdot\Ucal_c(t,t')\cdot \Jb(t')}}{(2s+1)(1-z)^2z^{2s+1}}. \nonumber\\
&& \label{eq:appinteval}
\end{eqnarray}
It is a simple matter now to combine the expression in Eq.~(\ref{eq:appinteval}) with the prefactor in Eq.~(\ref{eq:applong}) to obtain the expression for $G_s[\Jb]$.  After some algebra, we finally get
\begin{eqnarray}
G_s[\Jb] &=& \oint\frac{\mathrm{d}z}{2\pi i}\frac{\mathrm{e}^{\int_0^{\tau}\int_0^{\tau}\mathrm{d}t\,\mathrm{d}t'\, \Jb^{\star}(t)\cdot \Gcal_c(z,t,t')\cdot \Jb(t')}}{(2s+1)(1-z)^2z^{2s+1}}, \label{eq:appgsfin}
\end{eqnarray}
where $\Gcal_c(z,t,t')$ (which plays a role similar to that of a \emph{Green's function\/}) is given by the expression
\begin{eqnarray}
\Gcal_c(z,t,t') &=& \left(\frac{z}{1-z} +\Theta(t-t')  \right) \Ucal_c(t,t'). \label{eq:appgreen}
\end{eqnarray}
The quantity $\Theta(t-t')$ denotes the Heaviside (unit step) function, defined as
\begin{equation}
\Theta(t-t') = \left\{ \begin{array}{cc} 
                        1 & \mathrm{for}\,\,t>t' \\
                         0& \mathrm{for}\,\, t<t'
                         \end{array} \right. .
\end{equation}

It is straightforward to evaluate the integral over $z$ in Eq.~(\ref{eq:appgsfin}) if we wish, but the expression for $G_s[\Jb]$ is much more useful for doing calculations as is, and for this reason this is the expression used in the main Paper (see Eq.~(\ref{eq:gjbeval})). The (continuous) variable $z$ can be thought of as a \emph{conjugate variable} to the (discrete) variable $s$, and the right hand side of Eq.~(\ref{eq:appgsfin}) can be interpreted as an \emph{integral transform\/} between the $z$-representation and the $s$-representation, i.e. 
\begin{eqnarray}  
G_s[\Jb] &=& \oint \frac{\mathrm{d}z}{2\pi i}\frac{\widetilde{G}_z[\Jb]}{(2s+1)(1-z)^2z^{2s+1}} 
\end{eqnarray}
where we have
\begin{eqnarray}
\widetilde{G}_z[\Jb] &\equiv& \mathrm{e}^{\int_0^{\tau}\int_0^{\tau}\mathrm{d}t\,\mathrm{d}t'\, \Jb^{\star}(t)\cdot \Gcal_c(z,t,t')\cdot \Jb(t')}. 
\end{eqnarray}
In practice, it is easier to compute quantities using the continuous $z$-representation generating functional $\tilde{G}_z[\Jb]$ (where the expression is a simple Gaussian) and take the transform back into the $s$-representation as a final step. In the next subsection of this Appendix, we use $G_s[\Jb]$ to calculate the fidelity amplitude $\Acal_s$.  

%%%%%%%%%%%%%%%%%%%%%%%%%%%%%%%%%%%%%%%%%%
\subsection{Using $G_s[\Jb]$ to calculate the fidelity amplitude $\Acal_s$}
With the expression for $G_s[\Jb]$ (see Eq.~(\ref{eq:appgsfin})), we can make use of Eq.~(\ref{eq:appacalder}) to calculate $\Acal_s$ by taking functional derivatives of $G_s[\Jb]$. Because $G_s[\Jb]$ is given by Gaussian, the right hand side of Eq.~(\ref{eq:appacalder}) can be rewritten in an entirely equivalent representation by way of introducing \emph{fictitious\/} two component quantum fields $\boldsymbol\phi(t)^{\dag}\equiv (\phi_1^{\star}(t), \, \phi_2^{\star}(t))$ that are completely described in terms of the expectation value
\begin{equation}
\begin{array}{c}
\langle \boldsymbol\phi(t)\rangle_q = \langle \boldsymbol\phi^{\star}(t)\rangle_q = {\bf 0} \\
\langle \boldsymbol\phi(t) \boldsymbol\phi^{\dag}(t)\rangle_q = \Gcal_c(z,t,t') , \label{eq:appphi}
\end{array}
\end{equation} 
where the Green's function $\Gcal_c(z,t,t')$ is the same one given in Eq.~(\ref{eq:appgreen}), and $\langle \cdot \rangle_q$ denotes a quantum expectation value over the fields $\boldsymbol\phi(t)$ with all higher order expectation values are entirely determined through the use of Wick's theorem.  With the aid of Eq.~(\ref{eq:appphi}), it is straightforward to show that the expression (suppressing time arguments for brevity)
\begin{eqnarray}
\Acal_s &=& \oint\frac{\mathrm{d}z}{2 \pi i}\frac{\left\langle\mathrm{e}^{-i\epsilon\int_0^{\tau}\mathrm{d}t\, \boldsymbol\phi^{\dag}\Hcal_n\boldsymbol\phi}  \right\rangle_q}{(2s+1)(1-z)^2z^{2s+1}} \label{eq:appacalsexp}
\end{eqnarray}
is equivalent to the defining relation for $\Acal_s$ given in Eq.~(\ref{eq:appacalder}).  This formulation, given in terms of the dynamics of the fictitious fields $\boldsymbol\phi$, has the advantage of being physically more intuitive. 

We shall now evaluate the quantum expectation value by making use of the following cumulant expansion
\begin{eqnarray}
\left\langle\mathrm{e}^{-i\epsilon\int_0^{\tau}\mathrm{d}t\, \boldsymbol\phi^{\dag}\Hcal_n\boldsymbol\phi}  \right\rangle_q &=& \exp\left\{  \sum_{m=1}^{\infty}\Xcal_m\right\}, \nonumber\\
&& \label{eq:appcum}
\end{eqnarray}  
where the quantities $\Xcal_m$ are defined by the relation
\begin{eqnarray}
\Xcal_m &\equiv&\frac{1}{m!}\left\langle \!\!\left\langle \left( -i\epsilon\int_0^{\tau}\mathrm{d}t\, \boldsymbol\phi^{\dag}\Hcal_n\boldsymbol\phi \right)^m \right\rangle \!\! \right\rangle _q. \label{eq:appxi}
\end{eqnarray}
The double brackets $\langle\!\langle \cdot \rangle\!\rangle_q$ on the right hand side of Eq.~(\ref{eq:appxi}) denote cumulant averages.  It is a well known fact that in a diagrammatic expansion only \emph{connected\/} diagrams contribute to cumulant averages. In what follows, we shall use the diagrammatic expansion to  \emph{exactly} evaluate the right hand side of Eq.~(\ref{eq:appcum}), i.e. we will perform an exact resummation of all connected diagrams.

Given the form of Eq.~(\ref{eq:appcum}), we see that each diagram contributing to $\Xcal_m$ contains \emph{exactly} $m$ vertices and $m$ propagators.  Considering this together with the fact that only connected diagrams contribute (since we are taking cumulant averages), places a severe restriction on the form allowed for the diagrams: all contributing diagrams have the topology of a \emph{single} non-selfcrossing \emph{closed loop}.  For convenience, we assign a definite orientation to this loop by placing arrows on the propagators (but note that different orientations are not distinct and should not be counted as such).

\begin{figure}
	\includegraphics[scale=0.75]{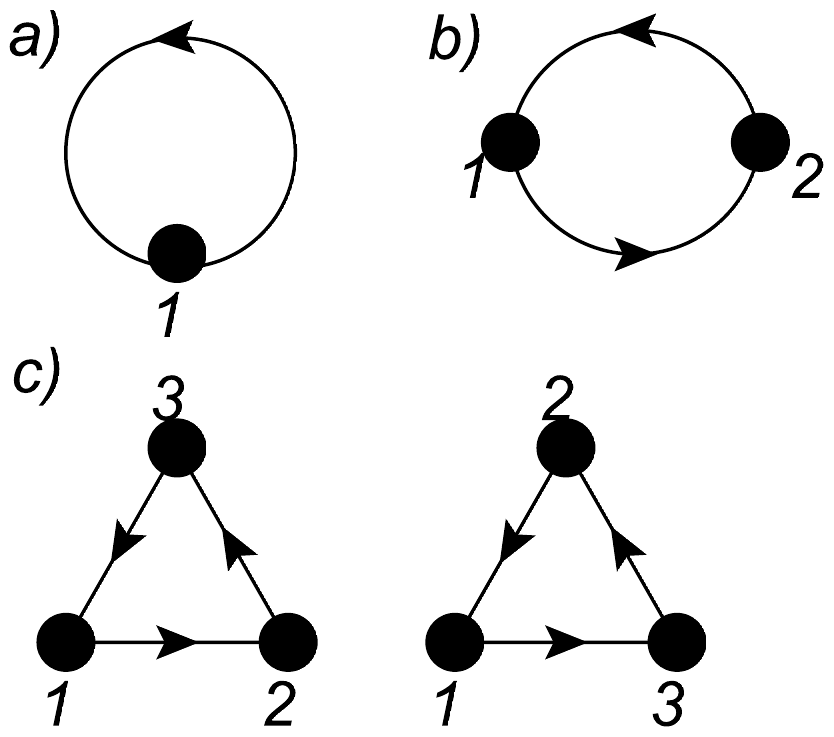}
		\caption{$a)$  Diagram corresponding to entire contribution for $\Xcal_1$. $b)$  Diagram corresponding to entire contribution to $\Xcal_2$.  $c)$  The $2!$ diagrams corresponding to the entire contribution for $\Xcal_3$. Both diagrams give the same contribution, since all variables are internal, so in practice one only need evaluate a single representative diagram (see discussion in text).  Likewise, for a given order $m$, all $(m-1)!$ diagrams give the exact same contribution. The expressions corresponding to the diagrams given in $a)$, $b)$, and $c)$ are given in Eqs.~(\ref{eq:appxcal1},\ref{eq:appxcal2},\ref{eq:appxcal3}) respectively (see text for Feynman rules).  For all diagrams, the integer labels $i=1,2,\cdots$ etc. next to the points refer to the time labels $t_i$, which are all dummy variables to be integrated over. Though we have chosen a counterclockwise orientation for the diagrams, one is free to choose whichever orientation one wishes (i.e. diagrams with different orientations are the exact same diagram). \label{fig:feynxcal12} }
\end{figure}

The Feynman rules are easily determined , and we state them here. In a given diagram, each vertex (with its time label $t_i$) corresponds to a factor $\Hcal_n(t_i)$, and each propagator with starting point $t_i$ and ending point $t_j$ (as determined from the direction of the arrow) corresponds to the Green's function $\Gcal_c(z,t_i,t_j)$. To obtain the value of an $m$th order diagram (i.e. one contributing to $\Xcal_m$), start at an arbitrary vertex and write down all of the factors corresponding to each vertex and propagator in the order determined by the direction of the arrows (taking care not to count the starting point twice). Then simply integrate over all time variables, take the matrix trace, and multiply by an overall prefactor $(-i\epsilon)^m/m$. That this prefactor is $(m-1)!$ times larger than expected from Eq.~(\ref{eq:appxi}) simply comes from that fact that for a given order $m$ there are $(m-1)!$ diagrams which differ only in their labels. Since all labels are dummy variables, all $(m-1)!$ diagrams give the same numerical contribution, hence the overall prefactor. To obtain the entire contribution to $\Xcal_m$, it suffices to evaluate a single representative diagram, so, i.e., in Fig.~(\ref{fig:feynxcal12}c) we only need to evaluate either the left or right diagram, but not both. 

%with the exception of the prefactor which we take to be$(-i\epsilon)^m/m$ to  account for all $(m-1)!$ diagrams at $m$th order.    

% It is easy to see that for a given order $m$ there are $(m-1)!$ diagrams which differ only in their labels.  Since all labels are dummy variables, all diagrams give the same numerical contribution.  To obtain the entire contribution to $\Xcal_m$, it then suffices to evaluate a single representative diagram using the same rules as above, with the exception of the prefactor which we take to be$(-i\epsilon)^m/m$ to  account for all $(m-1)!$ diagrams at $m$th order.    

The diagrams corresponding to $\Xcal_m$ for $m=1,2,3$ are given in Figs.~(\ref{fig:feynxcal12}a,b,c) respectively.  The integer labels displayed ($i=1,2,\cdots$ etc.) correspond the time labels $t_i$.  We have arbitrarily chosen to use a counterclockwise orientation when drawing the loop diagrams, though as we noted earlier different orientations correspond the the same diagram. As we have stated earlier, both diagrams shown in Fig.~(\ref{fig:feynxcal12}c) give the same contribution, since all labels correspond to internal variables.  As an example, we evaluate the diagrams shown in Fig.~(\ref{fig:feynxcal12}), giving expressions corresponding to $\Xcal_1$, $\Xcal_2$, and $\Xcal_3$.  We find
\begin{subequations}
\begin{eqnarray}
\Xcal_1&=& \frac{(-i\epsilon)^1}{1}\int_0^{\tau}\mathrm{d}t_1 \, \mathrm{Tr} \,\Hcal_n(t_1)\Gcal_c(z,t_1,t_1)  \label{eq:appxcal1} \\
\Xcal_2&=&\frac{(-i\epsilon)^2}{2}\int_0^{\tau}\mathrm{d}t_1\int_0^{\tau}\mathrm{d}t_2 \, \mathrm{Tr} \,\Big[ \Hcal_n(t_1)\Gcal_c(z,t_1,t_2) \nonumber\\
&& \times \Hcal_n(t_2)\Gcal_c(z,t_2,t_1) \Big] \label{eq:appxcal2} \\
\Xcal_3&=& \frac{(-i\epsilon)^3}{3}\int_0^{\tau}\mathrm{d}t_1\int_0^{\tau}\mathrm{d}t_2\int_0^{\tau}\mathrm{d}t_3 \, \mathrm{Tr}\, \Big[ \Hcal_n(t_1)   \nonumber\\
&&\times\Gcal_c(z,t_1,t_2) \Hcal_n(t_2)\Gcal_c(z,t_2,t_3) \Hcal_n(t_3)\Gcal_c(z,t_3,t_1) \Big], \nonumber\\
&& \label{eq:appxcal3} 
\end{eqnarray}
\end{subequations}
and it is straightforward to see to how the expression generalizes for arbitrary $\Xcal_m$.  

Now that we know the expression for all $\Xcal_m$, all that remains is to evaluate the sum $\sum_{m=1}^{\infty}\Xcal_m$.   Let us first introduce the shorthand notation
\begin{eqnarray}
\zeta &\equiv& \frac{z}{1-z}\nonumber\\
\Theta_{ij} &\equiv& \Theta(t_i-t_j),
\end{eqnarray}
and rewrite the expression for $\Xcal_m$ as follows: 
\begin{widetext}
\begin{eqnarray}
\Xcal_m &=& \frac{(-i\epsilon)^m}{m} \int_0^\tau\mathrm{d}t_1\int_0^\tau\mathrm{d}t_2\cdots \int_0^{\tau}\mathrm{d}t_m\; \mathrm{Tr}\left[\Hcal_n(t_1)\Gcal_c(z,t_1,t_2)\Hcal_n(t_2)\Gcal_c(z,t_2,t_3) \cdots \Hcal_n(t_m)\Gcal_c(z,t_m,t_1)  \right]\nonumber\\
&=&  \frac{(-i\epsilon)^m}{m} \int_0^\tau\mathrm{d}t_1\int_0^\tau\mathrm{d}t_2\cdots \int_0^{\tau}\mathrm{d}t_m\; \mathrm{Tr}\left[ \Hcal_n(t_1)\Ucal_c(t_1,t_2)\Hcal_n(t_2)\Ucal_c(t_2,t_3) \cdots \Hcal_n(t_m)\Ucal_c(t_m,t_1) \right] \nonumber\\
&&\times(\zeta+\Theta_{12})(\zeta+\Theta_{23})\cdots(\zeta+\Theta_{m1}) \nonumber\\
&=&  \frac{(-i\epsilon)^m}{m} \int_0^\tau\mathrm{d}t_1\int_0^\tau\mathrm{d}t_2\cdots \int_0^{\tau}\mathrm{d}t_m\; \mathrm{Tr}\left[ (\Ucal_c^{\dag}(t_1,0)\Hcal_n(t_1)\Ucal_c(t_1,0)) \cdots (\Ucal_c^{\dag}(t_m,0)\Hcal_n(t_m)\Ucal_c(t_m,0)) \right] \nonumber\\
&&\times(\zeta+\Theta_{12})(\zeta+\Theta_{23})\cdots(\zeta+\Theta_{m1}) \nonumber\\
&=&  \frac{(-i\epsilon)^m}{m} \int_0^\tau\mathrm{d}t_1\cdots \int_0^{\tau}\mathrm{d}t_m\; \mathrm{Tr}\left[ \Hcalt_n(t_1) \cdots\Hcalt_n(t_m)\right] (\zeta+\Theta_{12})\cdots(\zeta+\Theta_{m1}), \label{eq:appxcaln}
\end{eqnarray}
\end{widetext}
where in the second line of Eq~(\ref{eq:appxcaln}) we simply use the definition of $\Gcal_c(z,t,t')$  (see Eq.~(\ref{eq:appgreen})), in the third line we use the property $\Ucal_c(t,t')=\Ucal_c(t,0)\Ucal_c(0,t')=\Ucal_c(t,0)\Ucal_c^{\dag}(t',0)$ along with the cyclic property of traces $\mathrm{Tr}\,AB\cdots YZ=\mathrm{Tr}\,ZAB\cdots Y$, and in the fourth line the quantity $\Hcalt_n(t)$ is defined as
\begin{eqnarray}
\Hcalt_n(t) &=& \Ucal_c(t,0)^{\dag}\Hcal_n(t)\Ucal_c(t,0).
\end{eqnarray}
Note that $\Hcalt_n(t)$ is simply the noise Hamiltonian in the rotating frame, or equivalently, in the interaction picture (with respect to the control Hamiltonian $\Hcal_c(t,0)$).

At first sight, it may seem that the sum $\sum_m \Xcal_m$ cannot be explicitly carried out since the expression for $\Xcal_m$, as given in Eq.~(\ref{eq:appxcaln}), does not factorize (the factors of $\Theta_{ij}$ make it impossible to factorize the time integrals).  This situation can be remedied by rearranging $\sum_m\Xcal_m$ into a power series in $\zeta$
\begin{eqnarray}
\sum_{m=1}^{\infty}\Xcal_m &=& \sum_{m=1}^{\infty}\Ycal_m \zeta^m. \label{eq:appxy}
\end{eqnarray}
Unlike $\Xcal_m$, the coefficients $\Ycal_m$ \emph{do} factorize,  making this rearrangement advantageous. They take the form
\begin{eqnarray}
\Ycal_m &=& \mathrm{Tr} \, \frac{\Delta^m}{m}, \label{eq:appycal}
\end{eqnarray} 
where $\Delta$ is a $2\times2$ matrix to be determined below. With the expression in Eq.~(\ref{eq:appycal}), we can evaluate the sum in  Eq.~(\ref{eq:appxy}) and obtain the result
\begin{eqnarray}
\sum_{m=1}^{\infty}\Xcal_m &=& \mathrm{Tr}\, \log (\eins-\zeta\Delta)^{-1}, \label{eq:appcumsum}
\end{eqnarray} 
where $\eins$ is the $2\times 2$ unit matrix.

We will now show that Eq.~(\ref{eq:appycal}) is true, finding the matrix $\Delta$ in the process.  We note that an \emph{infinite set} of $\Xcal_q$ contribute to $\Ycal_m$, since \emph{all} $\Xcal_q$ with $q\geq m$ contain terms proportional to $\zeta^m$.  Starting with the first term, $\Ycal_0$ (i.e. the term in the sum $\sum_m\Xcal_m$ proportional to $\zeta^0$), we find that it exactly vanishes. This is due to the fact that the contribution coming from each $\Xcal_m$ is proportional to $\Theta_{12}\Theta_{23}\cdots\Theta_{m1}$, and in order for this to be nonvanishing we need $t_1>t_2\cdots>t_{m-1}>t_m>t_1$, an impossibility.  

Next, we seek the expression for $\Ycal_1$ by collecting all terms linear in $\zeta$ from $\sum_m\Xcal_m$. It is easy to see from Eq.~(\ref{eq:appxcaln}) that a given $\Xcal_m$ contributes exactly $m$ terms, furthermore all of these terms give the exact same contribution since they only differ in the labeling of internal variables.  It then suffices to take a single representative and multiply the result my $m$. Let us take as the representative the term proportional to $\Theta_{12}\Theta_{23}\cdots\Theta_{m-1,m}$. The effect of this factor is to simply cut off the limits in the time integrals so that $t_1>t_2\cdots>t_m$, and we are left with a \emph{time ordered} sequence of factors $\Hcalt_{n}(t_1)\Hcalt_{n}(t_2)\cdots\Hcalt_n(t_m)$ which can be easily factorized via the help of the time ordering operator $\Tcal$. The (linear in $\zeta$) contribution from $\Xcal_m$ is given explicitly by
\begin{eqnarray}
\Xcal_m &\to& \zeta (-i\epsilon)^m \mathrm{Tr}\,\int_0^\tau\mathrm{d}t_1\int_0^{t_1}\mathrm{d}t_2\cdots\int_0^{t_{m-1}}\mathrm{d}t_m \, \Hcalt_n(t_1)\nonumber\\
&&\times\Hcalt_n(t_2)\cdots\Hcalt_n(t_m) \nonumber\\
&=& \zeta\mathrm{Tr}\, \frac{1}{m!} \mathcal{T}\left[-i\epsilon\int_0^{\tau}\mathrm{d}t\;\Hcalt_1(t) \right]^m.
\end{eqnarray}
Collecting all terms in $\sum_m\Xcal_m$ proportional to $\zeta$, we obtain
\begin{eqnarray}
\Ycal_1 &=& \mathrm{Tr}\,\mathcal{T}\left\{\sum_{m=1}^{\infty}\frac{1}{m!} \left[-i\epsilon\int_0^{\tau}\mathrm{d}t\;\Hcalt_n(t) \right]^m \right\}\nonumber\\
&=& \mathrm{Tr}\Big[\mathcal{T}\mathrm{e}^{-i\epsilon\int_0^{\tau}\mathrm{d}t\;\Hcalt_1(t)}-\eins\Big] \nonumber\\
&\equiv& \mathrm{Tr}\,\Delta, \label{eq:appycal1}
\end{eqnarray}
giving us the sought for expression for the matrix $\Delta$.

Now we proceed to find $\Ycal_2$, showing that the general result given in Eq.~(\ref{eq:appycal}) holds. Looking at Eq.~(\ref{eq:appxcaln}), we see that all $\Xcal_m$ for $m\geq 2$ contain a term proportional to  $\zeta^2$ and therefore contribute to $\Ycal_2$. In order to more easily understand the pattern that emerges, let us consider the contributions from the first few $\Xcal_m$. The lowest order term to contribute to $\Ycal_2$ is $\Xcal_2$, from which we get the expression 
\begin{eqnarray}
\Xcal_2 &\rightarrow& \zeta^2 \frac{(-i\epsilon)^2}{2}\mathrm{Tr}\,\int_0^{\tau}\mathrm{d}t_1 \int_0^{\tau}\mathrm{d}t_2\; \Hcalt_n(t_1)\Hcalt_n(t_2) \nonumber\\
&=& \frac{\zeta^2}{2}\mathrm{Tr}\left[\frac{1}{1!}\mathcal{T}\left(-i\epsilon\int_0^{\tau}\mathrm{d}t \;\Hcalt_n(t) \right) \right. \nonumber\\
&&\left. \times \frac{1}{1!}\mathcal{T}\left(-i\epsilon \int_0^{\tau}\mathrm{d}t \;\Hcalt_n(t)\right)    \right], \label{eq:appxcal2zeta2}
\end{eqnarray}
where in the second line of Eq~(\ref{eq:appxcal2zeta2}) we have just rewritten the expression in a way that suggests what the pattern will be for higher order terms.

In order to more clearly see the pattern that emerges, let us work out the $\zeta^2$ order contribution coming from $\Xcal_3$. We have
\begin{widetext}
\begin{eqnarray}
\Xcal_3 &\rightarrow& \zeta^2 \frac{(-i\epsilon)^3}{3}\mathrm{Tr}\left[\int_0^{\tau}\mathrm{d}t_1\int_0^{\tau}\mathrm{d}t_2\int_0^{\tau}\mathrm{d}t_3\; \Hcalt_n(t_1)\Hcalt_n(t_2)\Hcalt_n(t_3) \left(\Theta_{12}+\Theta_{23}+\Theta_{31}  \right)     \right] \nonumber\\
&=& \zeta^2(-i\epsilon)^3 \mathrm{Tr}\left[\int_0^{\tau}\mathrm{d}t_1\int_0^{\tau}\mathrm{d}t_2\int_0^{\tau}\mathrm{d}t_3\; \Hcalt_n(t_1)\Hcalt_n(t_2)\Hcalt_n(t_3) \; \Theta_{12}     \right] \nonumber\\
&=& \zeta^2 (-i\epsilon)^3 \mathrm{Tr}\left[\left(\int_0^{\tau}\mathrm{d}t_1\int_0^{t_1}\mathrm{d}t_2\;\Hcalt_n(t_1)\Hcalt_n(t_2)\right)\left(\int_0^{\tau}\mathrm{d}t \Hcalt_n(t)   \right)    \right] \nonumber\\
&=& \frac{\zeta^2}{2}\mathrm{Tr} \left[\frac{1}{2!}\mathcal{T}\left(-i\epsilon\int_0^{\tau}\mathrm{d}t\; \Hcalt_n(t)\right)^2 \frac{1}{1!}\mathcal{T}\left(-i\epsilon\int_0^{\tau}\mathrm{d}t\; \Hcalt_n(t)   \right) + \frac{1}{1!}\mathcal{T}\left(-i\epsilon\int_0^{\tau}\mathrm{d}t\; \Hcalt_n(t)   \right)\frac{1}{2!}\mathcal{T}\left(-i\epsilon\int_0^{\tau}\mathrm{d}t\; \Hcalt_n(t)\right)^2 \right]\nonumber\\
&=& \frac{\zeta^2}{2}\, \mathrm{Tr}\, \sum_{j=1}^2\frac{1}{j!}\Tcal\left(-i\epsilon\int_0^{\tau}\mathrm{d}t\, \Hcalt_n(t) \right)^j \frac{1}{(2-j)!}\Tcal\left(-i\epsilon\int_0^{\tau}\mathrm{d}t \,\Hcalt_n(t)  \right)^{2-j}  .\nonumber\\
&&\label{eq:appxcal3zeta2}
\end{eqnarray}
\end{widetext}
In the second line we have used the fact that the terms proportional to $\Theta_{23}$ and $\Theta_{31}$ are identical to the one propotional to $\Theta_{12}$.  This can be seen by simply relabeling variables along with taking cyclic permutations of $\Hcalt_n(t_i)$, since the trace operation remains invariant under this.  In the third line, we simply apply $\Theta_{12}$ to cut off the integral over $t_2$, and note that the expression factorizes as the parentheses suggest.  In the third line we rewrite the expression as a sum of two terms, via the use of time-ordering operators $\mathcal{T}$, and by once again taking advantage of the invariance of the trace operation under cyclic permutations of matrices.  In the final line, we rewrite the sum is a way that is suggestive of how higher order $\Xcal_m$ contribute to $\Ycal_2$.

As one can easily guess from the expression in Eq.~(\ref{eq:appxcal3zeta2}) (we do not show the proof explicitly here, though it is easy to prove by induction) the contribution to $\Ycal_2$ coming from $\Xcal_m$ for general $m$ is given by the expression
\begin{eqnarray}
\Xcal_m &\rightarrow& \frac{\zeta^2}{2} \, \mathrm{Tr}\, \Big[\sum_{j=1}^{m-1}\frac{1}{j!}\mathcal{T}\left( -i\epsilon\int_0^{\tau}\mathrm{d}t\, \Hcalt_n(t) \right)^j \nonumber\\
&& \times \frac{1}{(m-j)!}\mathcal{T}\left( -i\epsilon\int_0^{\tau}\mathrm{d}t\, \Hcalt_n(t)  \right)^{m-j} \Big].
\end{eqnarray}
We are now ready to write down the entire contribution to $\Ycal_2$, coming from all $\Xcal_m$, where we find
\begin{eqnarray}
\Ycal_2 &=& \frac{1}{2}\sum_{m=2}^{\infty}\mathrm{Tr}\Big[\sum_{j=1}^{m-1}\frac{1}{j!}\mathcal{T}\left(-i\epsilon\int_0^{\tau}\mathrm{d}t\, \Hcalt_n(t)  \right)^j \nonumber\\
&& \times \frac{1}{(m-j)!}\mathcal{T}\left( -i\epsilon\int_0^{\tau}\mathrm{d}t\, \Hcalt_n(t)   \right)^{m-j} \Big] \nonumber\\
&=& \mathrm{Tr}\left[\frac{1}{2}\left(\mathcal{T}\mathrm{e}^{-i\epsilon\int_0^{\tau}\mathrm{d}t\, \Hcalt_n(t)  }-\eins      \right)^2 \right] \nonumber\\
&=& \mathrm{Tr}\, \frac{\Delta^2}{2}, \label{eq:appycal2}
\end{eqnarray}
where we see that in the last line of Eq.~(\ref{eq:appycal2}), the matrix $\Delta$ is the same matrix appearing in Eq.~(\ref{eq:appycal1}).

Continuing with the same procedure used above to determine $\Ycal_1$ and $\Ycal_2$, it is simple to show (though we do not explicitly show the proof here) that for general order $m$ we get the expression
\begin{eqnarray}
\Ycal_m &=& \mathrm{Tr} \, \frac{\Delta^m}{m},
\end{eqnarray}  
reproducing the expression given in Eq.~(\ref{eq:appycal}), which is what we intended to show. Using this expression, along with Eqs.~(\ref{eq:appxy}) and (\ref{eq:appcumsum}), we obtain the exact result for the quantum average in Eq.~(\ref{eq:appcum}).  We have
\begin{eqnarray}
\left\langle\mathrm{e}^{-i\epsilon\int_0^{\tau}\mathrm{d}t\, \boldsymbol\phi^{\dag}\Hcal_n\boldsymbol\phi}  \right\rangle_q &=&  \exp\left\{  \sum_{m=1}^{\infty}\Ycal_m \zeta^m \right\} \nonumber\\
&=& \exp\left\{\mathrm{Tr}\, \log (\eins-\zeta\Delta)^{-1}   \right\} \nonumber\\
&=& \exp \left\{\log \, \mathrm{Det}(\eins-\zeta\Delta)^{-1}    \right\} \nonumber\\
&=& \frac{1}{\mathrm{Det}(\eins -\zeta\Delta)}. \label{eq:appexpavg}
\end{eqnarray}  
With this expression, all that remains is to evaluate the integral over $z$ in Eq.~(\ref{eq:appacalsexp}) in order to obtain the expression for $\Acal_s$.  

Before doing so, let us first define the matrix $\delta$
\begin{eqnarray}
\delta &\equiv& \eins + \Delta \nonumber\\
&=& \Tcal \mathrm{e}^{-i\epsilon \int_0^{\tau}\mathrm{d}t \, \Hcalt_n(t)}, \label{eq:appdeltadef}
\end{eqnarray}
in terms of which the analysis to follow simplifies.  Recalling that $\zeta\equiv z/(1-z)$, we see that
\begin{eqnarray}
\frac{1}{\mathrm{Det}(\eins -\zeta\Delta)} &=& \frac{(1-z)^2}{\mathrm{Det}(\eins-z\delta)}, \label{eq:appdetdet}
\end{eqnarray}
where we used the fact that for any $2\times2$ matrix $A$ and c-number $c$ we have $\mathrm{Det}(cA)=c^2\mathrm{Det}\, A$.  This is a very convenient representation since the factor of $(1-z)^2$ in Eq.~(\ref{eq:appdetdet}) cancels a similar factor appearing in Eq.~(\ref{eq:appacalsexp}) which simplifies the evaluation of the integral over $z$.  It also turns out to be very convenient to reexpress the determinant on the right hand side of Eq.~(\ref{eq:appdetdet}) as 
\begin{eqnarray}
\Big[\mathrm{Det}(\eins-z\delta)\Big]^{-1} &=& \Big[1-z\,\mathrm{Tr}\,\delta +z^2\,\mathrm{Det}\, \delta \Big]^{-1}\nonumber\\
&=& \Big[1-z\,\mathrm{Tr}\, \delta+z^2 \Big]^{-1}, \label{eq:appdetexpand}
\end{eqnarray}
where the first line is an \emph{exact\/} relation for \emph{any\/} $2\times2$ matrix $\delta$, and in the second line we used the fact that $\delta$ is a $SU(2)$ matrix (see Eq.~(\ref{eq:appdeltadef})) with unit determinant. 

As a final step before evaluating the integral over $z$ in Eq.~(\ref{eq:appacalsexp}) to find $\Acal_s$, we take advantage of the fact that the right hand side of Eq.~(\ref{eq:appdetexpand}) takes the form of a \emph{generating function\/} for the \emph{Chebyshev polynomials of the second kind\/}~\cite{modernanalysis}, which play an important role in the development of spherical harmonics in four dimensions. We have
\begin{eqnarray}
\frac{1}{1-z\, \mathrm{Tr}\,\delta+z^2} &=& \sum_{j=0}^{\infty} V_j\Big(\frac{1}{2}\,\mathrm{Tr}\,\delta \Big)\, z^j, \label{eq:appcheby}
\end{eqnarray}
where $V_j(x)$ denotes the $j$th order Chebyshev polynomial of the second kind. Eq.~(\ref{eq:appcheby}) is valid as long as the conditions $|z|<1$ and $|\frac{1}{2}\,\mathrm{Tr}\,\delta|\leq1$ apply, which is always the case for our problem. Using Eqs.~(\ref{eq:appdetdet}), (\ref{eq:appdetexpand}), and (\ref{eq:appcheby}) we find a very useful representation for Eq.~(\ref{eq:appexpavg}):
\begin{eqnarray}
\left\langle\mathrm{e}^{-i\epsilon\int_0^{\tau}\mathrm{d}t\, \boldsymbol\phi^{\dag}\Hcal_n\boldsymbol\phi}  \right\rangle_q &=& (1-z)^2\sum_{j=0}^{\infty}V_j\Big(\frac{1}{2}\,\mathrm{Tr}\,\delta  \Big) \, z^j. \nonumber\\
&& \label{eq:appexpconv}
\end{eqnarray} 

We are now ready to evaluate Eq.~(\ref{eq:appacalsexp}) to find $\Acal_s$.  Using Eq.~(\ref{eq:appexpconv}), we find
\begin{eqnarray}
\Acal_s &=& \oint\frac{\mathrm{d}z}{2\pi i}\frac{\left\langle\mathrm{e}^{-i\epsilon\int_0^{\tau}\mathrm{d}t\, \boldsymbol\phi^{\dag}\Hcal_n\boldsymbol\phi}  \right\rangle_q}{(2s+1)(1-z)^2z^{2s+1}} \nonumber\\
&=& \frac{1}{2s+1}\sum_{j=0}^{\infty}\oint\frac{\mathrm{d}z}{2\pi i}\frac{V_j\Big(\frac{1}{2}\,\mathrm{Tr}\,\delta \Big)}{z^{2s+1-j}} \nonumber\\
&=& \frac{1}{2s+1}V_{2s}\Big(\frac{1}{2}\, \mathrm{Tr}\, \delta   \Big),
\end{eqnarray}  
where in going from the second to the third line, we use Cauchy's residue theorem.
In order to arrive at the expression for $A_s$ shown in Eq.~(\ref{eq:acali}) in the main Paper, we make use of the following relation
\begin{eqnarray}
V_j(\cos\, \theta) &=& \frac{\sin \, (j+1)\theta}{\sin \,\theta}.
\end{eqnarray} 
We then have
\begin{eqnarray}
\Acal_s &=& \frac{1}{2s+1} \frac{\sin\Big[(2s+1)\cos^{-1}\Big(\frac{1}{2}\,\mathrm{Tr}\,\delta \Big)\Big]}{\sin\Big[\cos^{-1}\Big(\frac{1}{2}\,\mathrm{Tr}\,\delta \Big)\Big]} \nonumber\\
&=& \frac{1}{2s+1}\sum_{j=-s}^{s}\mathrm{e}^{-2ji \cos^{-1}\Big(\frac{1}{2}\, \mathrm{Tr}\,\delta  \Big)}, \label{eq:appalmfin}
\end{eqnarray}
where the expression given in the second line turns out to be more convenient for calculations. By carrying out the sum in the second line, one arrives at the expression given in the right hand side of the first line, showing that the two expressions are entirely equivalent.  Taking the simplest case, $s=1/2$, we get the expression
\begin{eqnarray}
\Acal_{1/2} &=& \frac{1}{2}\Big[\mathrm{e}^{i\cos^{-1}\Big(\frac{1}{2}\,\mathrm{Tr}\,\delta \Big)}+\mathrm{e}^{-i\cos^{-1}  \Big(\frac{1}{2}\,\mathrm{Tr}\,\delta \Big)}   \Big] \nonumber\\
&=& \frac{1}{2}\, \mathrm{Tr}\, \delta \nonumber\\
&=& \frac{1}{2}\Tcal \mathrm{e}^{-i\epsilon\int_0^{\tau}\mathrm{d}t\,\Hcalt_n(t)}, \label{eq:appahalf}
\end{eqnarray}
where in the third line we made use of Eq.~(\ref{eq:appdeltadef}).  Making use of Eq.~(\ref{eq:appahalf}), we finally arrive at the expression given in Eq.~(\ref{eq:acali}) in the main Paper:
\begin{eqnarray}
\Acal_s &=&  \frac{1}{2s+1}\sum_{j=-s}^{s}\mathrm{e}^{-2ji \cos^{-1}\Acal_{1/2}}.
\end{eqnarray}
This relation is remarkable in the fact that it essentially shows us we can understand the behavior of the spin $s$ system entirely in terms of quantities associated with the spin $1/2$ system.  The dynamics of a spin $s$ system is entirely contained in the dynamics of a spin-half system.

%%%%%%%%%%%%%%%%%%%%%%%%%%%%%%%%%%%%%%%%%%%%%%%%%

\bibliography{spinrefs}

\end{document}